\documentclass[12pt,preprint]{aastex}

\begin{document}
 \title{The evolution of primordial binary open star clusters: 
        mergers, shredded secondaries and separated twins} 
 \shorttitle{Primordial binary open clusters}

 \author{R.~de~la~Fuente~Marcos and C.~de~la~Fuente~Marcos}
  \affil{Suffolk University Madrid Campus, C/ Vi\~na 3,
         E-28003, Madrid, Spain}
  \email{raul@galaxy.suffolk.es}

  \begin{abstract}
     The basic properties of the candidate binary star cluster 
     population in the Magellanic Clouds and Milky Way are similar.
     The fraction of candidate binary systems is $\sim$10\% and 
     the pair separation histogram exhibits a bimodal distribution 
     commonly attributed to the transient nature of these clusters. 
     However, if primordial pairs cannot survive for long as 
     recognizable bound systems, how are they ending up? Here, we 
     use simulations to confirm that merging, extreme tidal 
     distortion and ionization are possible depending on the 
     initial orbital elements and mass ratio of the cluster pair. 
     The nature of the dominant evolutionary path largely depends 
     on the strength of the local galactic tidal field. Merging is 
     observed for initially close primordial binary clusters but 
     also for wider pairs in nearly parabolic orbits. Its 
     characteristic timescale depends on the initial orbital 
     semi-major axis, eccentricity, and cluster pair mass ratio, 
     becoming shorter for closer, more eccentric equal mass pairs. 
     Shredding or extreme tidal distortion of the less massive 
     cluster and subsequent separation is observed in all pairs 
     with appreciably different masses. Wide pairs steadily evolve 
     into the separated twins state characterized by the presence 
     of tidal bridges and cluster separations of 200-500 pc after 
     one Galactic orbit. In the Galaxy, the vast majority of 
     observed binary system candidates appear to be following this 
     evolutionary path which translates into the dominant peak 
     (25-30 pc) in the observed pair separation distribution. The 
     secondary peak at smaller separations (10-15 pc) can be 
     explained as due to close pairs in almost circular orbits 
     and/or undergoing merging. Merged clusters exhibit both 
     peculiar radial density and velocity dispersion profiles 
     shaped by synchronization and gravogyro instabilities. Both 
     simulations and observations show that, for the range of open 
     cluster parameters studied here, long term binary cluster 
     stability in the Milky Way disk is highly unlikely. 
  \end{abstract}


  \section{Introduction}
     Observations show that the fraction of potential binary open clusters is 
     not negligible. This finding is not surprising as open clusters are born in 
     star complexes (Efremov 1978, 1995, 2010) with several of these objects 
     being formed nearly at the same time and in close proximity (for a recent 
     review on this topic see, e.g., de la Fuente Marcos \& de la Fuente Marcos 
     2008, 2009a). An early attempt to study this topic by Rozhavskii et al. 
     (1976) concluded that the fraction of multiple systems among open clusters 
     was $\sim$20\%. Almost two decades later and using a larger sample, 
     Subramaniam et al. (1995) found that about 8\% of open clusters may be 
     genuine binaries. Loktin (1997) further strengthened the idea that multiple 
     open clusters are not uncommon by providing a catalogue of 31 probable 
     multiple systems. 

     The question of the possible existence of a sizeable fraction of candidate 
     binary clusters in the Galactic disk has been revisited again by de la 
     Fuente Marcos \& de la Fuente Marcos (2009b). Using complete, 
     volume-limited samples, they have found consistent and statistically robust 
     figures: at the Solar Circle, at least 12\% of all open clusters could be 
     part of potential binary systems. This number agrees well with the ones 
     previously found by various authors for the Magellanic Clouds (see Dieball 
     et al. 2002 for details). The cluster binary fraction is not only 
     numerically similar; both in the Clouds and Milky Way the pair separation 
     histogram shows a conspicuous bimodal distribution. Yet another common 
     characteristic is the practical absence of close, almost coeval pairs among 
     objects older than about 100 Myr. All this body of solid observational 
     evidence strongly indicates that, if primordial binary star clusters do 
     form, they appear not to be able to survive as such for long. Obvious 
     evolutionary paths include merging, tidal disruption and ionization but, 
     under what conditions merging is favored? what is the characteristic 
     merging timescale? what is the (dynamic/kinematic) signature of an open 
     cluster formed by merging? how does tidal disruption proceed? what happens 
     to ionized primordial pairs? is the initial number of stars in the cluster 
     a factor to consider in the final outcome? can we recover the initial 
     dynamical properties of primordial pairs? are long term stable binary 
     clusters possible at all?

     In this work we examine the evolution of simulated primordial binary 
     clusters in an attempt to provide convincing answers to the above 
     questions. This Paper is organized as follows: In \S 2 we present the 
     simulations. We show our results in \S 3. In \S 4 we discuss our results 
     and our conclusions are presented in \S 5. 

  \section{Simulations} 
     The goal of these simulations is to understand how primordial binary 
     clusters evolve, how they merge when they do, and how the resulting 
     merger remnants look like. We use the direct $N$-body code {\small 
     NBODY6.GPU} (Aarseth \& Nitadori 2009). This $N$-body code is an improved 
     parallelized version of the standard scalar code {\small NBODY6} (Aarseth 
     2003) for use with multi-core machines and Nvidia 
     CUDA\footnote{http://www.nvidia.com/cuda}-capable devices. This code has 
     been recently applied to study the dynamics of globular clusters (Heggie \& 
     Giersz 2009; Trenti et al. 2010) and it is publicly available from the IoA 
     web site \footnote{http://www.ast.cam.ac.uk/$\sim$sverre/web/pages/nbody.htm}.

     The initial conditions for each simulation (see details in Table 
     \ref{results}) were generated by randomly selecting stellar positions 
     and velocities according to the Plummer distribution (Plummer 1911). The 
     Plummer density profile is given by
     \begin{equation}
        \rho(r) = \frac{3 M}{4 \pi R_p^3} \frac{1}{(1 + (r/R_p)^2)^{5/2}} \,,
     \end{equation}
     where $r$ is the radial coordinate, $M$ is the total mass of the cluster 
     and $R_p$ is the Plummer radius, a scale parameter which sets the 
     characteristic size of the central regions of the cluster. It is related to 
     the half-mass radius by $R_h \simeq 1.305 R_p$ (Aarseth \& Fall 1980; 
     Heggie \& Hut 2003). For details on how to set up a numerical Plummer model 
     see Kroupa (2008), also Aarseth et al. (1974). In the simulations, the 
     actual input parameter that defines the initial physical size of the 
     cluster is the virial radius, given by
     \begin{equation}
       R_{vir} \ = \ - \frac{G M^{2}}{4 E}  \,,
     \end{equation}
     where $G$ is the gravitational constant and $E$ is the total energy of
     the system. For a Plummer model the potential energy is (e.g., Binney \& 
     Tremaine 2008)
     \begin{equation}
       W = -\frac{3 \pi G M^2}{32 R_p}\,,
     \end{equation}
     and $E$ = $W$/2. Therefore,
     \begin{equation}
       R_{vir} \ = \ \frac{16}{3 \pi} \ R_p  \,.
     \end{equation}
     Each simulation follows the evolution of two Plummer models in an initially 
     bound orbit. Input parameters are the initial separation (apoclustron 
     distance, $S_o$ = 10, 20, and 30 pc) and orbital eccentricity ($e_o$ = 0.0, 
     0.2, 0.4, 0.6, 0.8, and 0.95). The apoclustron is function of both semi-major 
     axis and eccentricity, $S_o = a_o (1 + e_o)$. Therefore and if we assume that
     the two clusters are point-like masses moving in Keplerian orbits, the initial 
     orbital period of the system can be approximated by
     \begin{equation}
        P_{orb} (\mbox{Myr}) = 94 \ \left(\frac{S_o}{1 + e_o}\right)^{3/2} \ 
                                     \frac{1}{\sqrt{M_1 + M_2}} \,,
                                      \label{period}
     \end{equation}
     where $M_1$ and $M_2$ are the cluster masses, the apoclustron is in pc
     and the masses in $M_{\odot}$ (see actual values in Table \ref{results}). 
     Stellar masses in the range [0.17, 10.0] $M_{\odot}$ are drawn from a 
     Salpeterian IMF with an average stellar mass of 0.5 $M_{\odot}$. Salpeter 
     (1955) used the observed luminosity function for the Solar Neighborhood and 
     theoretical evolution times to derive an initial mass function (IMF) which 
     may be approximated by a power-law: 
     \begin{equation}
       n(m) \propto m^{-\alpha} \,,  
     \end{equation}
     where $n(m)$ is the number of stars per unit mass interval. The value of 
     $\alpha$ is 2.35 for masses between 0.4 and 10.0 $M_{\odot}$. The IMF used
     (a single power-law) and the mass range of stars ([0.17, 10.0] $M_{\odot}$)
     are equivalent to the realistic canonical two-part power-law over the
     mass range [0.08, 10.0] $M_{\odot}$ as described in (e.g.) Kroupa (2008). 
     Neglecting stars more massive than 10 $M_{\odot}$ is a good approximation
     since such stars are rare (see, e.g., the $m_{max}$-star cluster mass relation
     of Weidner et al. 2010). All stars are started on a zero-age main sequence 
     with a uniform composition of hydrogen, $X$ = 0.7, helium, $Y$ = 0.28, and 
     metallicity, $Z$ = 0.02. Stellar evolution is computed according to the 
     algorithm described by Aarseth (2003). Primordial binaries were not included 
     but binary and multiple system formation was allowed and observed. External 
     perturbations were represented by a fixed galactic tidal field. The tidal 
     radius is given by the expression (e.g., Aarseth 2003)
     \begin{equation}
       R_{T} = \left(\frac{G M}{4 A (A - B)}\right)^{1/3} \,,
     \end{equation}
     where $A$ and $B$ are the Oort's constants of Galactic rotation. For a
     star located at that distance from the cluster, the central attraction of 
     the cluster is balanced by the Galactic tidal force. The Oort's constants 
     are chosen to be $A$ = 14.4, $B$ = -12.0 km s$^{-1}$ kpc$^{-1}$ (Binney \& 
     Tremaine 2008). The cluster pair is assumed to move in a circular orbit at 
     the Solar Circle with no passing molecular clouds (see Aarseth 2003, pp. 
     127-129, for additional details). No escaping stars were removed from the 
     calculations. 
 
     The simulations presented here have been performed on a Dell Precision 
     T5500 + Nvidia Tesla S1070 system. The T5500 has 2 quad-core Intel Xeon 
     E5540 processors at 2.53 GHz, the S1070\footnote{http://www.nvidia.com/tesla} 
     includes 4 GPU cores each with 240 stream processors running at 1.44 GHz 
     for a total of 960 processing units. The Tesla parallel computing 
     architecture is relatively new (Lindholm et al. 2008) but it is already 
     in use for astrophysical applications (e.g. Thompson et al. 2010; Jonsson 
     \& Primack 2010; Schive et al. 2010; Trenti et al. 2010).  

  \section{Results}
     Here we describe the main results for the four different sets of models: 18 
     models with two equal clusters ($N_1$ = $N_2$ = 4096, $M = N \ <m_{\star}>$, 
     $q = M_2/M_1 = N_2/N_1$, $q$ = 1), 36 models with $N_1$ = 4096, $N_2$ = 2048, 
     $q$ = 0.5, 18 models with $N_1$ = 4096, $N_2$ = 1024, $q$ = 0.25, and 4 
     additional control models for single clusters submitted to the same tidal 
     field but with $N$ = 8192, 6144, 5120, and 4096, respectively. Calculations 
     were stopped after a simulated time of 210 Myr or one Galactic rotation at 
     the Solar Circle; therefore, all the most interesting phases of the binary 
     cluster evolution could be studied. The state of the system at the end of the 
     simulation is summarized in Table \ref{results}. The merging timescale has 
     been computed by assuming that a merger takes place if the pair separation 
     becomes less than the average value of the core radius for observed open 
     clusters. The average value of core, half-mass, and tidal radii for real open 
     clusters are 1/2/10 pc (Binney \& Tremaine 2008), respectively. This rather 
     conservative merging criterion may appear arbitrary but it responds to what 
     is observed in our calculations: if the pair separation goes under 1 pc, the 
     two clusters quickly merge (see Figure \ref{trajectory}). From a strictly 
     observational point of view, star clusters separated by less than twice the 
     core radius are unlikely to be identified as separated objects; therefore,
     2 pc could also be a valid (and less conservative) observational criterion 
     for merging.  

     \subsection{$N_1 = N_2$}
\begin{figure}
        \epsscale{0.8}
        \plotone{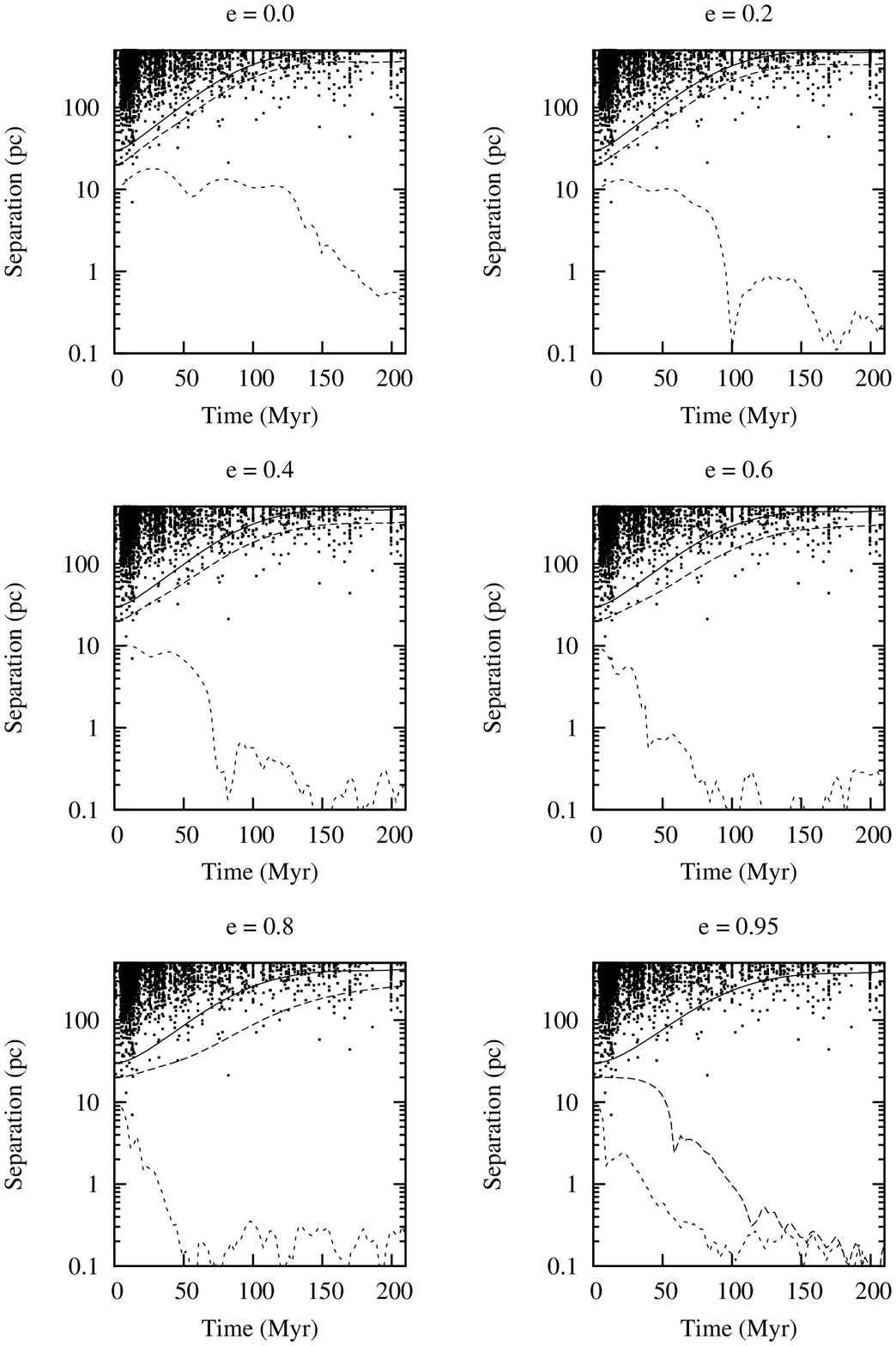} 
        \caption{Evolution of the orbital separation for models of two 
                 clusters with $N$ = 4096 for a total of 8192 stars. The 
                 three curves on each panel are for initial separations
                 (apoclustron distance) 10, 20, 30 pc and the value of 
                 the initial orbital 
                 eccentricity indicated on the panel label. For merged 
                 models, cluster centers were computed only using stars 
                 within 30 pc of the pair centers. The points correspond 
                 to actual open cluster pairs from the de la Fuente Marcos 
                 \& de la Fuente Marcos (2009b) study. The sample displayed
                 is made of pairs with age difference $<$ 30 Myr, separation 
                 $<$ 500 pc and age $<$ 210 Myr. The age of the pair is that 
                 of its younger member.
                 \label{evolequal}}
     \end{figure}

\begin{figure}
        \includegraphics[angle=270,scale=0.7]{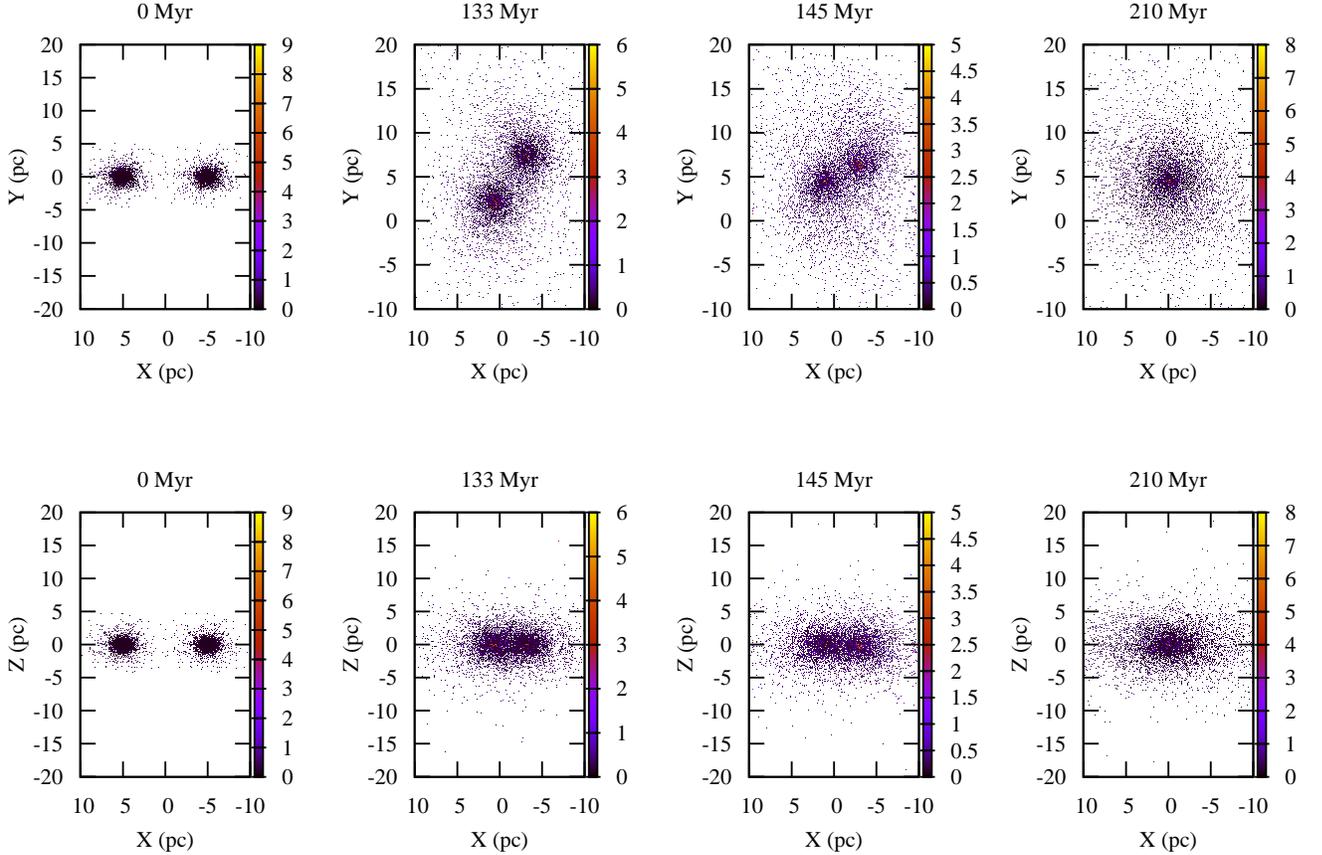} 
        \caption{Merging: representative snapshots in the $X-Y$ plane (top) 
                 and the $X-Z$ plane (bottom) of the evolution of the model 
                 with $q$ = 1.0, $S_o$ = 10.0 pc (initial separation, apoclustron 
                 distance), and $e_o$ = 0.0. Color is
                 related to the mass of the simulated star in $M_{\odot}$
                 according to the key provided. Merging for this model is the 
                 slowest of all the models ending in merging. Once the two 
                 clusters get closer than 2 pc merging proceeds very quickly. 
                 The $X$ axis points towards the galaxy center, $Y$ is tangent 
                 to the galactocentric pair motion, and $Z$ is perpendicular 
                 to the galaxy plane. For real open clusters, we do not have 
                 access to the $X-Y$ view. In the $X-Z$ plane, the merger
                 remnant looks significantly more elliptical than the 
                 single/individual cluster. By the end of the simulation all
                 the stars more massive than about 3.8 $M_{\odot}$ have already
                 evolved away from the main sequence. More massive objects
                 are all stellar remnants or (less frequently) stellar mergers.
                 No objects were removed from the calculations. The initial
                 orbital period was nearly 46 Myr. 
                 \label{merging}}
     \end{figure}

\begin{figure}
        \epsscale{0.8}
        \plotone{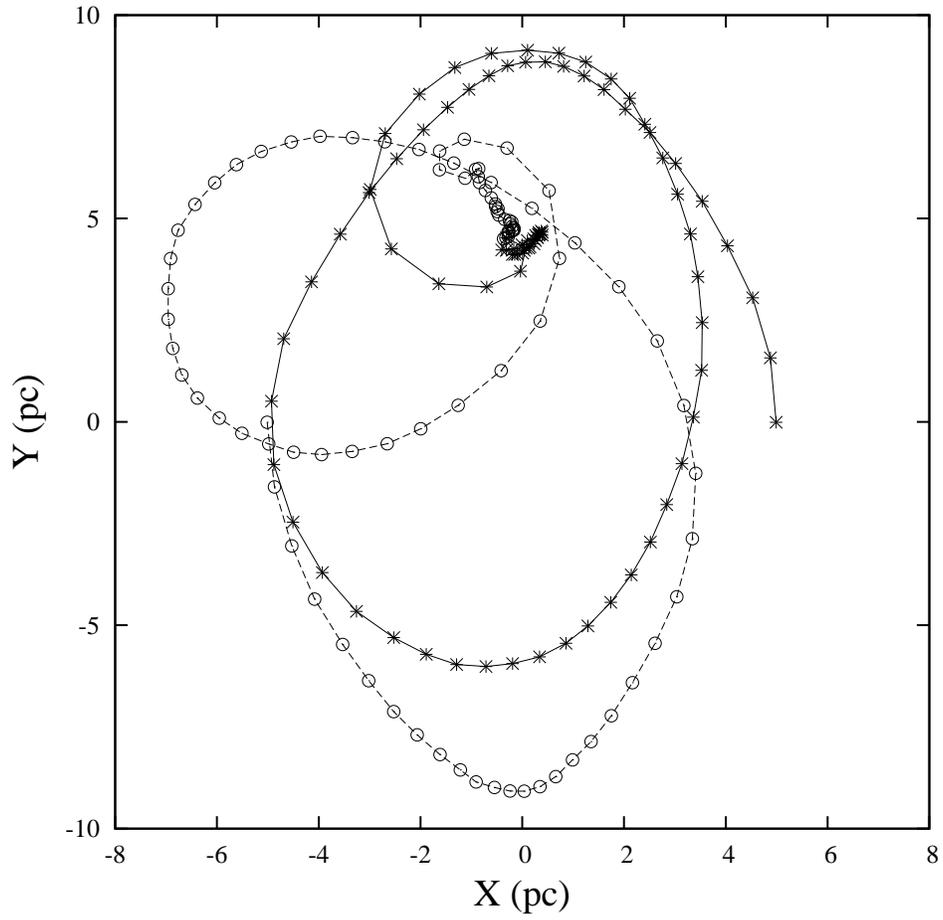} 
        \caption{The trajectory of the center of masses of the merging clusters
                 for the model displayed in Figure \ref{merging}. The 
                 time difference between consecutive points is 2.33 
                 Myr.
                 \label{trajectory}}
     \end{figure}

        Figure \ref{evolequal} shows the evolution of the orbital separation for 
        models with two equal orbiting clusters and different initial 
        apoclustron and eccentricity. The initial virial radius for the two 
        Plummer models is 1 pc with a mean density of 489 $M_{\odot}$ pc$^{-3}$. 
        Our numerical results show that two outcomes are possible: merging or a 
        special type of soft ionization. Long term binary stability is not 
        observed. Merging is found in all initially close pairs.
        Less eccentric pairs take longer to merge, with the circular case 
        taking almost four full initial orbital periods to complete the merging
        process. For wider pairs, a consistent behavior is also found; clusters 
        gradually increase their separation with the rate of increase slowing 
        down over time. This rate depends on the initial value of the eccentricity of
        the pair and it is slower for higher eccentricity. An example of merging 
        appears in Figures \ref{merging}, 
        \ref{trajectory} (model with two clusters in an initially circular orbit and 
        apoclustron 10 pc) and the outcome of ionization after 210 Myr is displayed 
        in Figure \ref{separatedtwins} (this figure is for a non-equal mass case but 
        the outcome is similar for the equal $N$ case). The ionized clusters that we 
        call "separated twins", after Theis (2001), are not completely unrelated 
        as a rather conspicuous tidal bridge is always observed. Formation of separated 
        twins is driven by mass loss induced by stellar evolution, two-body relaxation, 
        and (less important) mutual tidal disruption. In relative terms, 
        separated twins retain a proportionally higher fraction of stars, which 
        is not surprising: following Innanen et al. (1972), for two clusters 
        separated by a distance larger than three times the outer radius of each 
        cluster, the amount of mutual disruption is rather negligible. Mergers
        are less rich in relative terms, as they have lost a fraction of their 
        original stars due to mutual disruption and instabilities during the merging 
        process. The longer the merging process takes, the larger the amount of mutual 
        disruption. In the $X-Z$ plane, merger remnants appear significantly
        more elliptical than their isolated counterparts (see Figure \ref{merging}).
        The merger process imparts rotation to the resulting stellar system in
        the same sense as the orbital motion (Alladin et al. 1985) due to a tidal
        torque. 

     \subsection{$N_1 \neq N_2$}
        Young candidate binary open clusters appear to show a clear trend; in
        general, the two members of the pair have rather different radii 
        (de la Fuente Marcos \& de la Fuente Marcos 2009c). This 
        could be a property intrinsic to open cluster formation in complexes, the 
        result of dynamical interaction, or caused by an observational selection 
        effect. If intrinsic to open cluster formation, primordial binary clusters 
        may ordinarily sport a massive primary and a less massive secondary. On 
        the other hand, not having the same initial population is likely to have 
        an impact on the tidal evolution of the pair. This group of simulations 
        is aimed at studying this case. The primary cluster is as described in 
        the previous section. 

        \subsubsection{q = 0.5}
           Here we explore the evolution of primordial cluster pairs in which the
           secondary of the system has half the population of the primary, $N_{2} 
           = N_{1}/2$. Two different sub-cases are investigated: i) both clusters 
           have the same stellar density and ii) the secondary is half the size of 
           the primary, i.e. the secondary is denser than the primary.

\begin{figure}
        \epsscale{0.8}
        \plotone{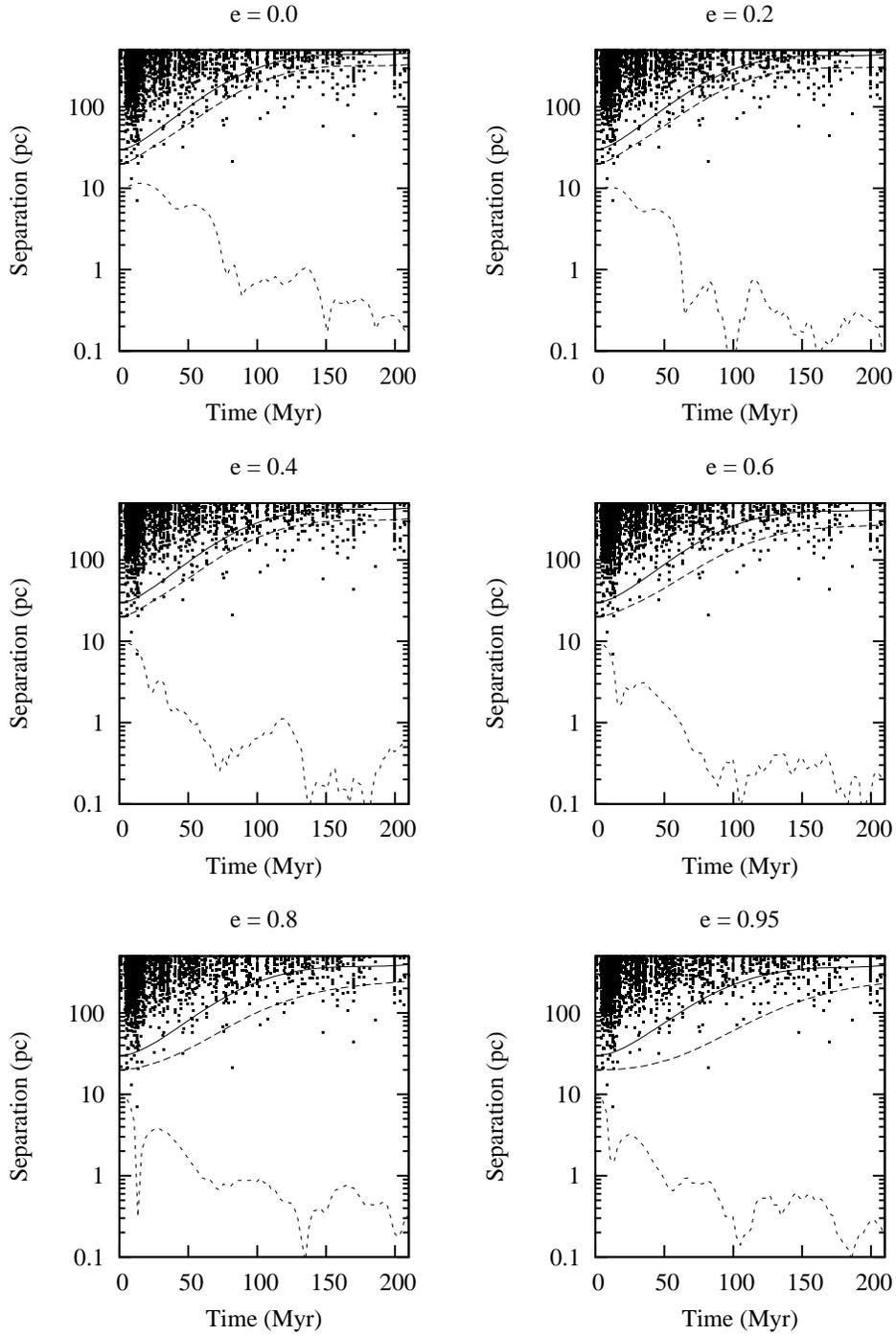} 
        \caption{Same as Figure \ref{evolequal} but for models with $q$ = 0.5
                 and the same average density.
                 \label{evolhalfsd}}
     \end{figure}

\begin{figure}
        \epsscale{0.8}
        \plotone{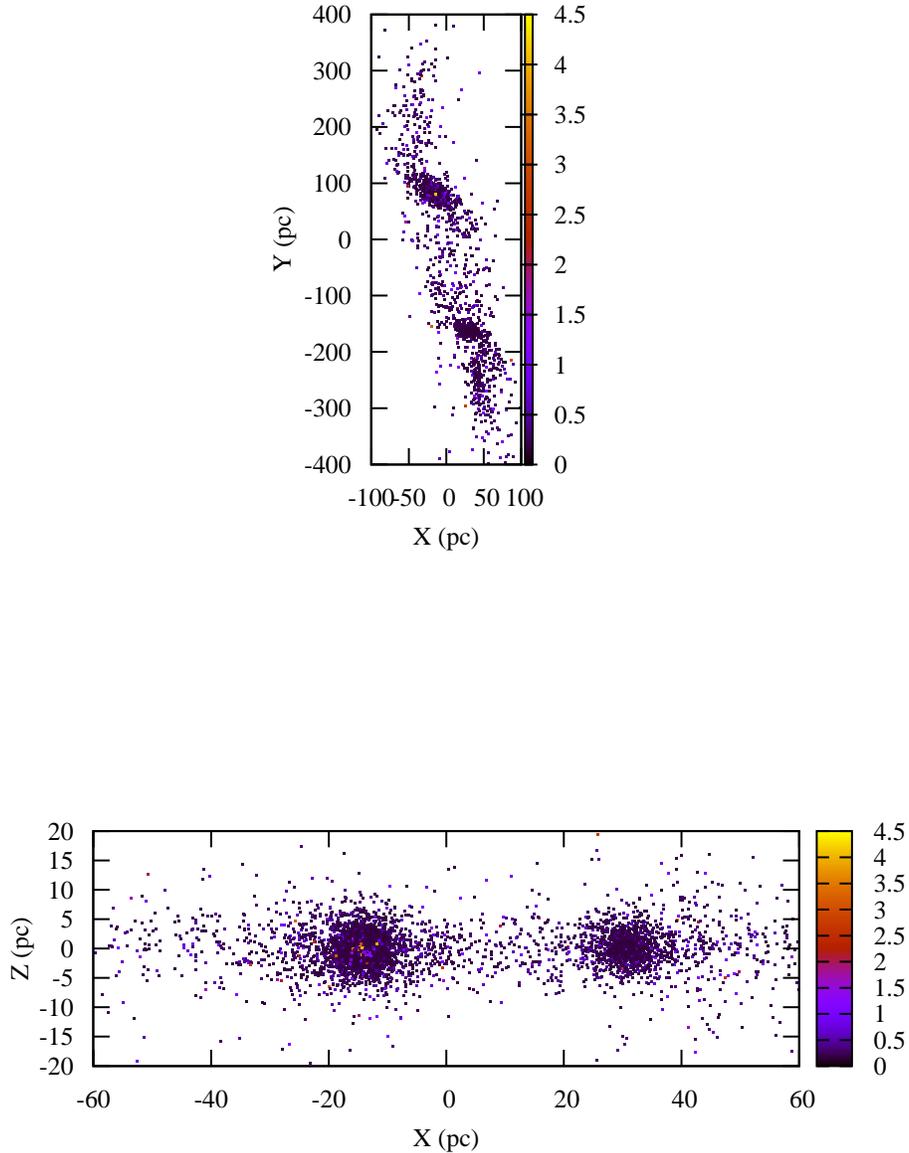} 
        \caption{Separated twins: representation in the $X-Y$ plane (top)
                 and the $X-Z$ plane (bottom) after 210 Myr for the model 
                 $q$ = 0.5 (equal average density), $S_o$ = 20.0 pc and 
                 $e_o$ = 0.95. Gradual separation is the slowest for this 
                 model. For other models following this evolutionary path
                 the main difference is in the larger separation at 210 
                 Myr. The famous Double Cluster ($h + \chi$ Persei pair, 
                 NGC 869/NGC 884) is a clear prototype for this pair type 
                 (actual physical separation $>$ 200 pc). Another obvious 
                 candidate is the pair NGC 659/NGC 663 (see de la Fuente 
                 Marcos \& de la Fuente Marcos 2009b for details). Colors
                 are as in Figure \ref{merging}.
                 \label{separatedtwins}}
     \end{figure}

           \subsubsubsection{i) same density}
              The secondary cluster has a virial radius of 0.79 pc so the mean 
              density is the same for both primary and secondary clusters. The 
              behavior of the orbital separation is similar to that found in the 
              previous section but the soft ionization process is slightly 
              slower (see Figure \ref{evolhalfsd}). The merging timescale is also 
              affected: mergers for models with $e_o >$ 0.5 take longer now but
              those for less eccentric pairs are faster. Merging is only observed 
              for models with an initial apoclustron distance of 10.0 pc. Cores of 
              merger remnants are more extended than those observed in the previous 
              case.

\begin{figure}
        \epsscale{0.8}
        \plotone{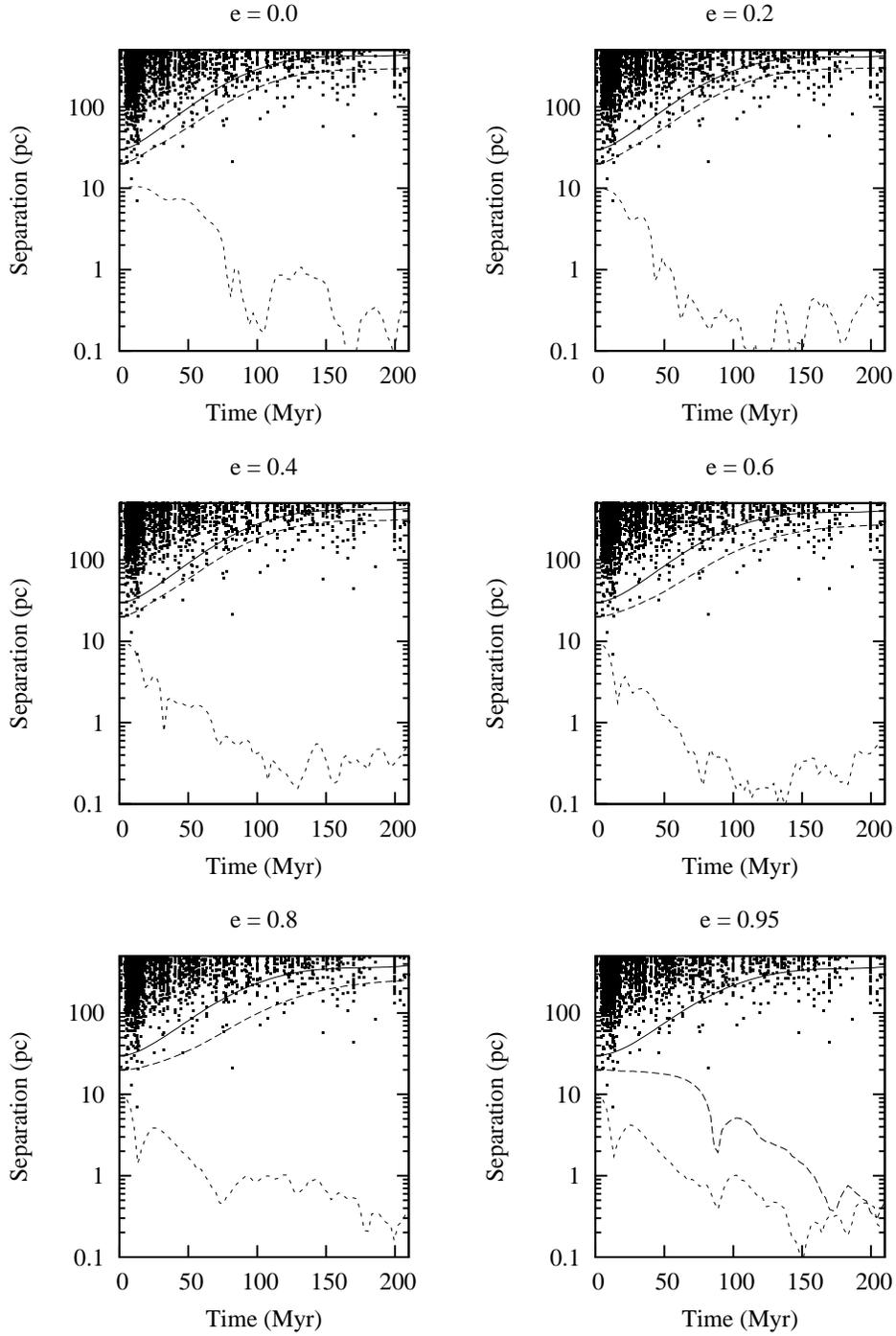} 
        \caption{Same as Figure \ref{evolhalfsd} but for models with $q$ = 0.5
                 and different average density (see the text for details).
                 Density appears to have a minor role on the overall evolution 
                 of the cluster pair.
                 \label{evolhalfdd}}
     \end{figure}

           \subsubsubsection{ii) different density}
              The secondary cluster has a virial radius of 0.5 pc with a mean 
              density of 1956 $M_{\odot}$ pc$^{-3}$. The behavior of the orbital 
              separation is similar to that found in the previous section (see 
              Figure \ref{evolhalfdd}). This is to be expected: a study by Sensui 
              et al. (2000) showed that the internal structure of galaxies does not
              play a role in the merging time-scales (inside a galaxy cluster) --
              only the distribution of galaxies inside the cluster matters. This
              argument should also hold for star clusters in a star cluster
              complex (Fellhauer et al. 2002, 2009). The single main difference 
              appears for highly eccentric models in which merging is observed in 
              two cases (as in $N_1$ = $N_2$): $S_o$ = 10.0 and 20.0 pc. 

        \subsubsection{q = 0.25}
\begin{figure}
        \epsscale{0.8}
        \plotone{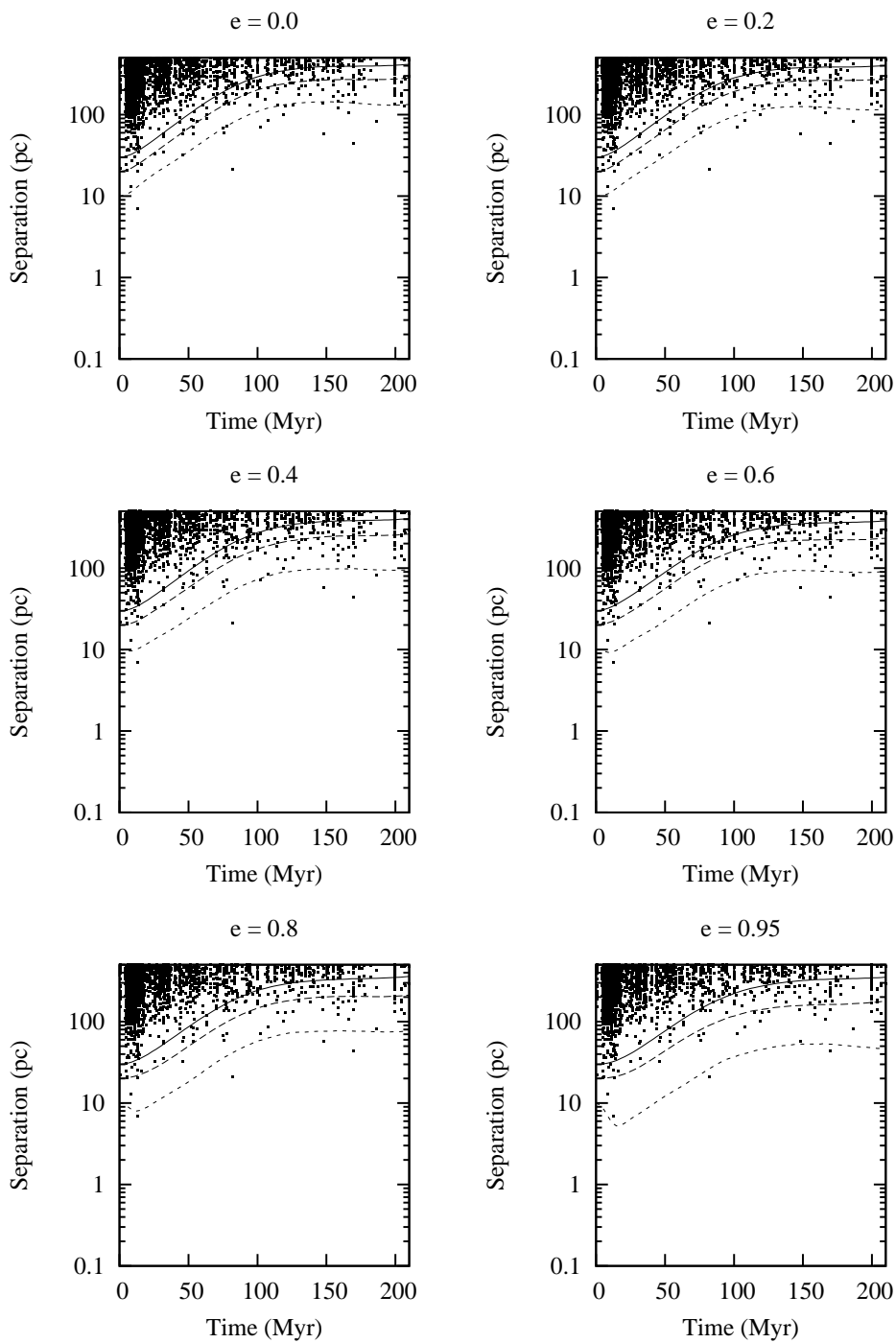} 
        \caption{Same as Figure \ref{evolequal} but for models with $q$ = 0.25
                 and the same average density. This result is different from 
                 that found by Portegies Zwart \& Rusli (2007) but their
                 simulations do not take into account the background galactic 
                 tidal field.
                 \label{evolfourth}}
     \end{figure}

\begin{figure}
        \includegraphics[angle=270,scale=0.7]{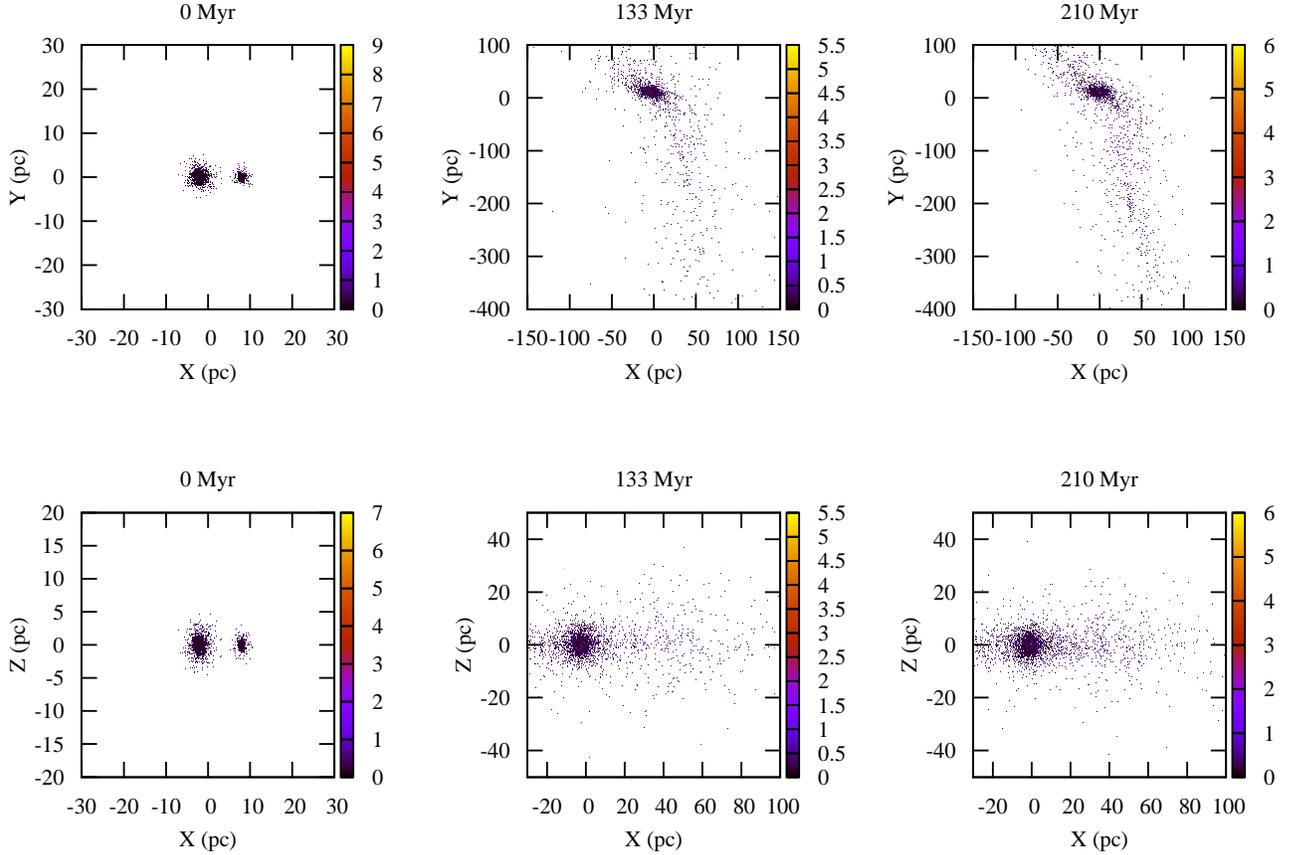} 
        \caption{Shredding of the secondary: representative snapshots in the $X-Y$ plane (top) 
                 and the $X-Z$ plane (bottom) of the evolution of the model 
                 with $q$ = 0.25, $S_o$ = 10.0 pc, and $e_o$ = 0.95. Colors
                 are like in Figure \ref{merging}. Extreme tidal distortion of the 
                 less massive cluster and subsequent separation is observed 
                 for all pairs with appreciably different mass ratio. The
                 identifiable cluster at 133, 210 Myr is the primary. The secondary
                 cluster gets shredded in a relatively short timescale (see Fig.
                 \ref{earlytidal}). 
                 The $X$ axis points towards the galaxy center, $Y$ is tangent 
                 to the galactocentric pair motion, and $Z$ is perpendicular 
                 to the galaxy plane. For real open clusters, we do not have 
                 access to the $X-Y$ view.  
                 \label{spaghetti}}
     \end{figure}

\begin{figure}
        \includegraphics[angle=270,scale=0.7]{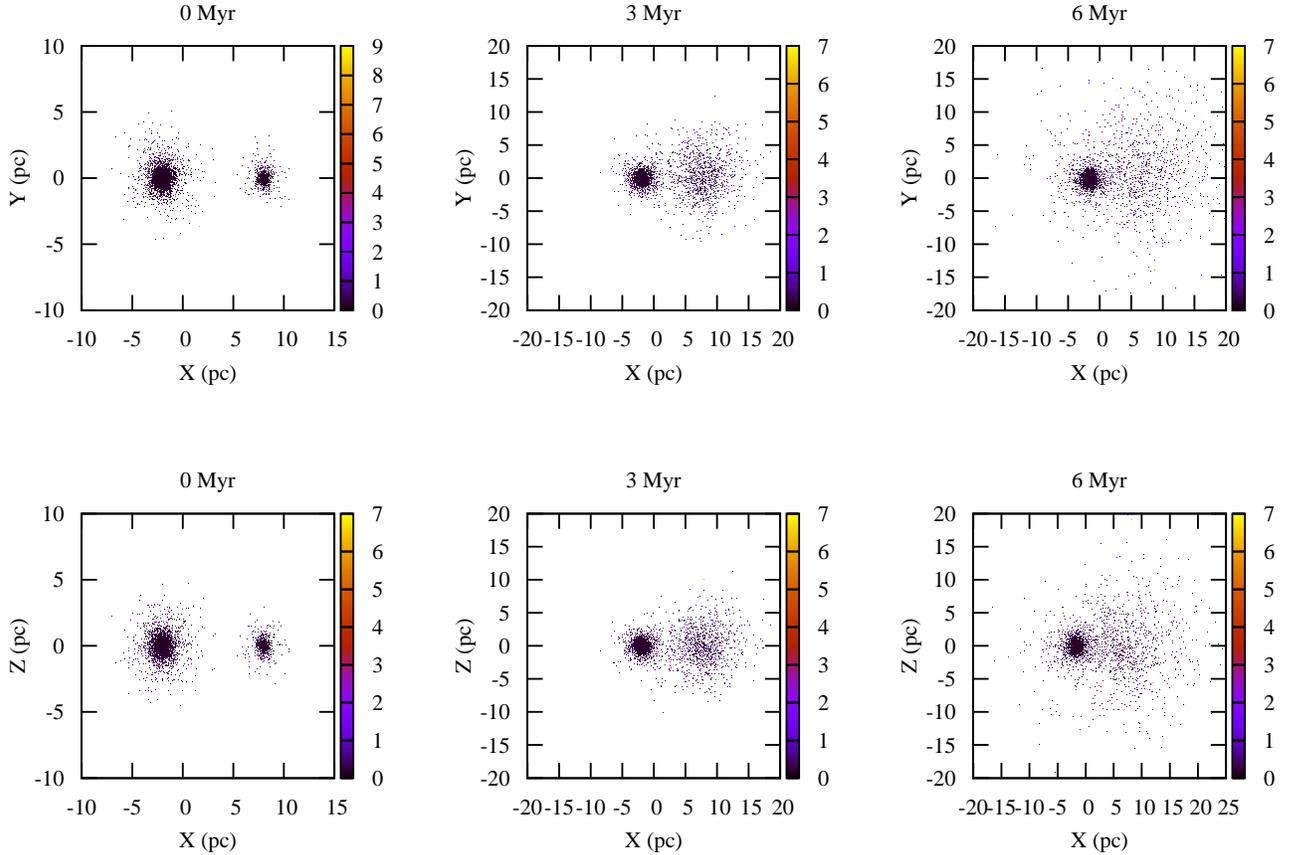} 
        \caption{Early evolution of the model displayed in Figure \ref{spaghetti}.
                 This behavior is similar to the one identified by Portegies Zwart
                 \& Rusli (2007): the initially less massive cluster expands 
                 quickly initiating mass transfer to the more massive cluster. Fast
                 expansion of the secondary cluster is the result of stellar
                 mass loss (see the text for details). 
                 As for observed open clusters, de la Fuente Marcos \& de la Fuente
                 Marcos (2009c) concluded that binary cluster candidates in the
                 Galactic disk appear to show a tendency to have components of
                 different physical size. Our results clearly indicate that this 
                 effect may well be the result of dynamical interactions.
                 \label{earlytidal}}
     \end{figure}

           This set of simulations is designed to study the impact of enhanced 
           tidal forces on the secondary cluster. With a virial radius of 0.63 
           pc, it still has the same reference mean density used 
           throughout this study. The dynamical behavior of the pair is now 
           substantially different (see Figure \ref{evolfourth}). The tidal 
           disruption timescale is much shorter than that for merging. Eventual 
           destruction of the less massive cluster is observed in all cases, 
           including close pairs. In some cases, the secondary cluster appears 
           extremely distorted and elongated with no clearly identifiable core 
           (see Figure \ref{spaghetti}): in other words, the secondary cluster 
           gets torn apart in a relatively short timescale. Technically 
           speaking, shredded clusters are different from typical open 
           cluster remnants, they look more like stellar streams and they may
           be rather young. Remnants of shredded clusters may show up in 
           kinematic studies even if they cannot be detected as stellar 
           overdensities. Tidal shredding is a form of shearing by differential
           rotation. The secondary cluster in proximity to the most massive
           primary becomes stretched out by tidal forces. The secondary 
           distends and flattens in the direction of the primary evolving as to
           minimize its gravitational potential energy becoming an ovoid stretched
           along the axis connecting the two bodies. Figure \ref{earlytidal} shows 
           the early evolution of the model displayed in Figure \ref{spaghetti}. 
           The secondary cluster undergoes early rapid expansion mainly due to stellar 
           mass loss. In addition, the characteristic timescale for equipartition
           of kinetic energy is much shorter for the secondary cluster contributing
           to its overall faster internal dynamical evolution. Coupling of these two 
           processes effectively drives the cluster away from equilibrium as the total 
           potential energy per unit mass quickly decreases. This behavior is, in 
           principle, similar to the one described in Portegies Zwart \& Rusli 
           (2007): the secondary expands quickly initiating mass transfer to the 
           primary. There is, however, a major difference: extreme tidal 
           distortion is not observed in their models (see comments below). 
           In our calculations and as the massive companion's gravitational
           pull is stronger on the overextended secondary cluster's near side
           than on the far side, the secondary cluster is literally shredded with
           the help of the galactic tidal field.

     \subsection{Control models}
        For the purpose of comparison, we computed four single Plummer models with 
        consistent tidal field: the virial radii were 1.26, 1.14, 1.08 and 1.00 for 
        a mean density of 489 $M_{\odot}$ pc$^{-3}$. The rest of the details of the
        simulation are common to the binary simulations. After 210 Myr the 
        values of the core, half-mass, and tidal radii were 0.095/1.92/18.0 pc ($N$ 
        = 8192), 0.106/2.23/18.0 pc ($N$ = 6144), 0.37/2.62/18.0 pc ($N$ = 5120) and
        0.34/2.61/18.0 pc ($N$ = 4096). The average values of the same magnitudes 
        for the merged clusters were: 0.29/4.17/22.6 pc ($q$ = 1.0) and 
        4.35/6.37/20.5 pc ($q$ = 0.5). The central regions of merger remnants are 
        significantly more extended than those of equivalent single models. Merger 
        remnants are less populated than an equivalent single cluster of the same 
        age as a result of mutual disruption of the pair during the merging process.
         
     \subsection{How realistic are these models?}
        Typical densities of candidate bound open clusters younger than 13 Myr are 
        in the range 4-400 $M_{\odot} \ pc^{-3}$ (Wolff et al. 2007) with the upper
        limit represented by $\chi$ and $h$ Persei. Slesnick et al. (2002) estimated
        that the total mass of the stars with $M > 0.1 M_{\odot}$ in $h$ and $\chi$
        Persei is 3700 and 2800 $M_{\odot}$, respectively, although these are,
        probably, lower limits for the masses of these clusters. Therefore, the 
        characteristics of the simulated clusters match well those of the most
        massive classical open clusters. These are the ones expected to be able to
        survive for several Galactic rotations and, therefore, statistically 
        speaking more likely to be observed. Their values are, however, far from those 
        typical of starburst clusters with densities in the range $10^{3}$ to several 
        10$^{5}$ $M_{\odot} \ pc^{-3}$ and $N$ = several 10$^4$-10$^5$ (see, e.g., 
        Pfalzner 2009). Similar trends have been found for star clusters in other 
        galaxies (Pfalzner \& Eckart 2009).

  \section{Discussion}
     The above results can only be properly understood within the context of the 
     Galactic tidal field; i.e. cluster tidal radii and separations play a major
     role in the outcome of primordial binary cluster evolution. The strength of
     the cluster-cluster interaction is maximum when the intercluster separation
     becomes smaller than the tidal radii. This interpretation was proposed in de 
     la Fuente Marcos \& de la Fuente Marcos (2009c) in the form of a 
     classification scheme (\ref{bincri}) based only in the values of separation 
     ($S$) and tidal radii ($R_{Ti}$, $i$ = 1, 2):
     \begin{equation}
        {\rm Cluster \ Pairs} \left\{ \begin{array}{l}
                                      {\rm Detached}, \ R_{T1} + R_{T2} < S \\
                                      \begin{array}{c}  
                                       {\rm Interacting} \\
                                       R_{T1} + R_{T2} > S
                                      \end{array}
                                       \left\{
                                        \begin{array}{l}
                                           {\rm Weak}, R_{T1} \ {\rm AND}
                                                       \ R_{T2} < S \\
                                           {\rm Semi-Detached}, \ R_{T1} 
                                                       \ {\rm OR} \ R_{T2} < S \\
                                           {\rm In-Contact}, \ R_{T1} 
                                                       \ {\rm AND} \ R_{T2} > S
                                        \end{array}
                                      \right.
                                    \end{array}
                            \right.
        \label{bincri}
     \end{equation} 
     The scheme was implemented by considering the available observational evidence. 
     For open cluster pairs of similar mass ratio in the detached and weakly 
     interacting categories, ionization into the separated twins state is the observed
     evolutionary path. Semi-detached and in-contact pairs merge in a timescale
     that depends strongly on the orbital eccentricity. Very eccentric pairs merge
     in a timescale of nearly 10 Myr and, therefore, the actual observation of the merging
     process may be very difficult as it may happen even before the embedded phase
     ends. If the mass ratio is appreciably different, extreme tidal distortion
     followed by actual destruction of the smallest cluster is always observed. The
     above interpretation has strong implications on what should be observed within the
     Milky Way at different Galactocentric distances and in other galaxies. Weaker
     galactic tidal fields increase the probability of observing semi-detached or
     in-contact cluster pairs and, eventually, mergers. The tidal force gradient 
     determines the cluster tidal radius (e.g. Elmegreen \& Hunter 2010). Even if the 
     fraction of primordial binary clusters may well be similar across different 
     galaxies, the preferential evolutionary path could be rather different: ionization 
     being dominant in massive galaxies and the central regions of galaxies in general.
      
     \subsection{How do these results compare with actual data?}
\begin{figure}
        \epsscale{0.8}
        \plotone{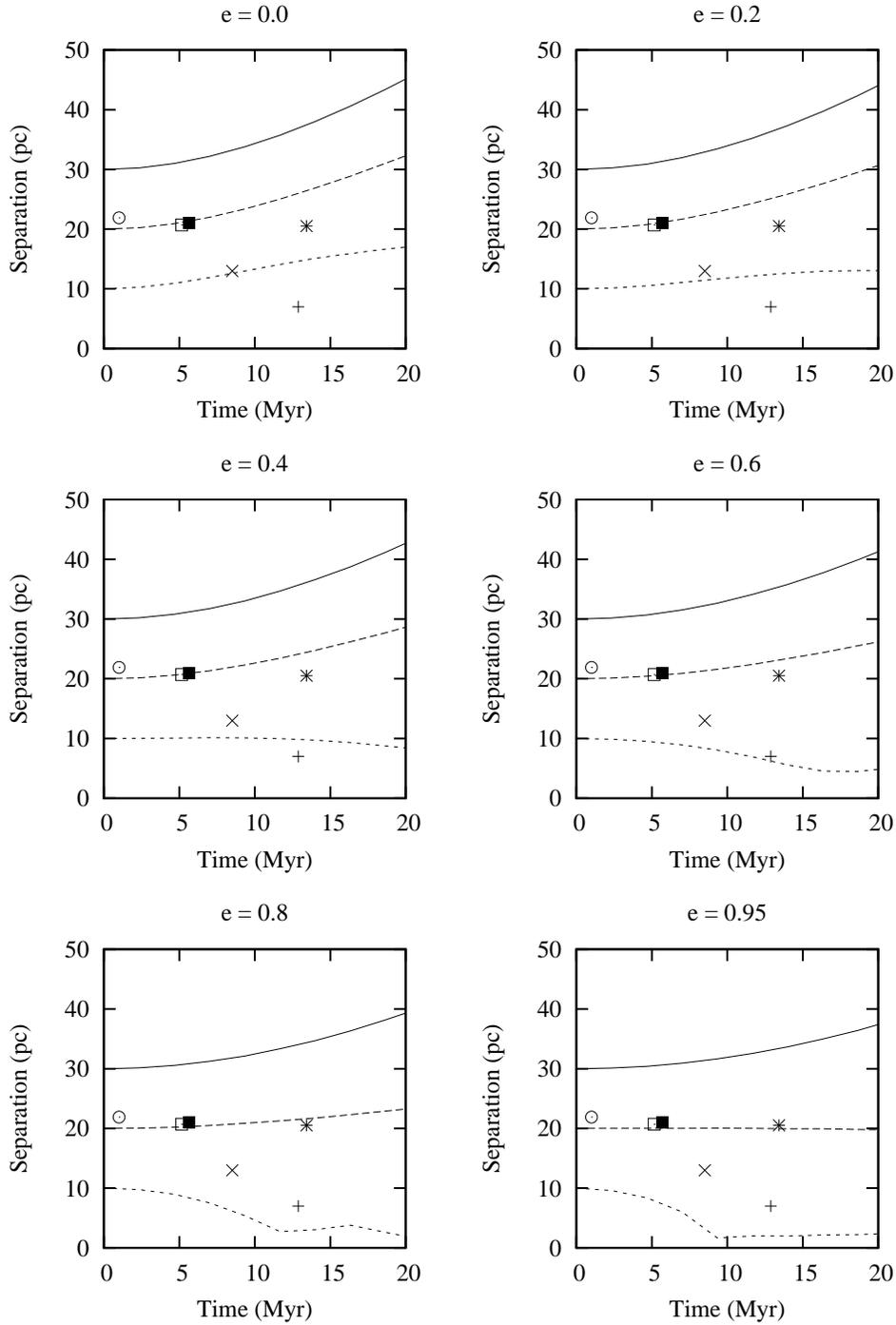} 
        \caption{Comparison between the time evolution of the pair 
                 separation in models of two clusters with $N$ = 4096 
                 (detail of Figure \ref{evolequal}) and real data for
                 strong candidates to be undergoing merging from de la Fuente Marcos 
                 \& de la Fuente Marcos (2009b) study. The sample displayed (symbol, age, separation, 
                 heliocentric distance) includes
                 NGC 1976/NGC 1981 ($+$, 13 Myr, 7 pc, 400 pc), ASCC 20/ASCC 16 ($\times$, 8 Myr, 13 pc, 460 pc), 
                 Collinder 197/ASCC 50 ($*$, 13 Myr, 20 pc, 838 pc), NGC 6250/Lynga 14 ($\Box$, 5 Myr, 21 pc,
                 865 pc), 
                 NGC 3324/NGC 3293 (filled $\Box$, 6 Myr, 21 pc, 2327 pc), and NGC 6613/NGC 6618 
                 ($\odot$, 1 Myr, 22 pc, 1296 pc). 
                 The cluster pair Trumpler 22/NGC 5617 (not shown in this plot) may also be in this category
                 (82 Myr, 21 pc, 1516 pc). The age difference of all these pairs is $<$ 30 Myr.
                 \label{realmerg}}
     \end{figure}

\begin{figure}
        \epsscale{0.3}
        \plotone{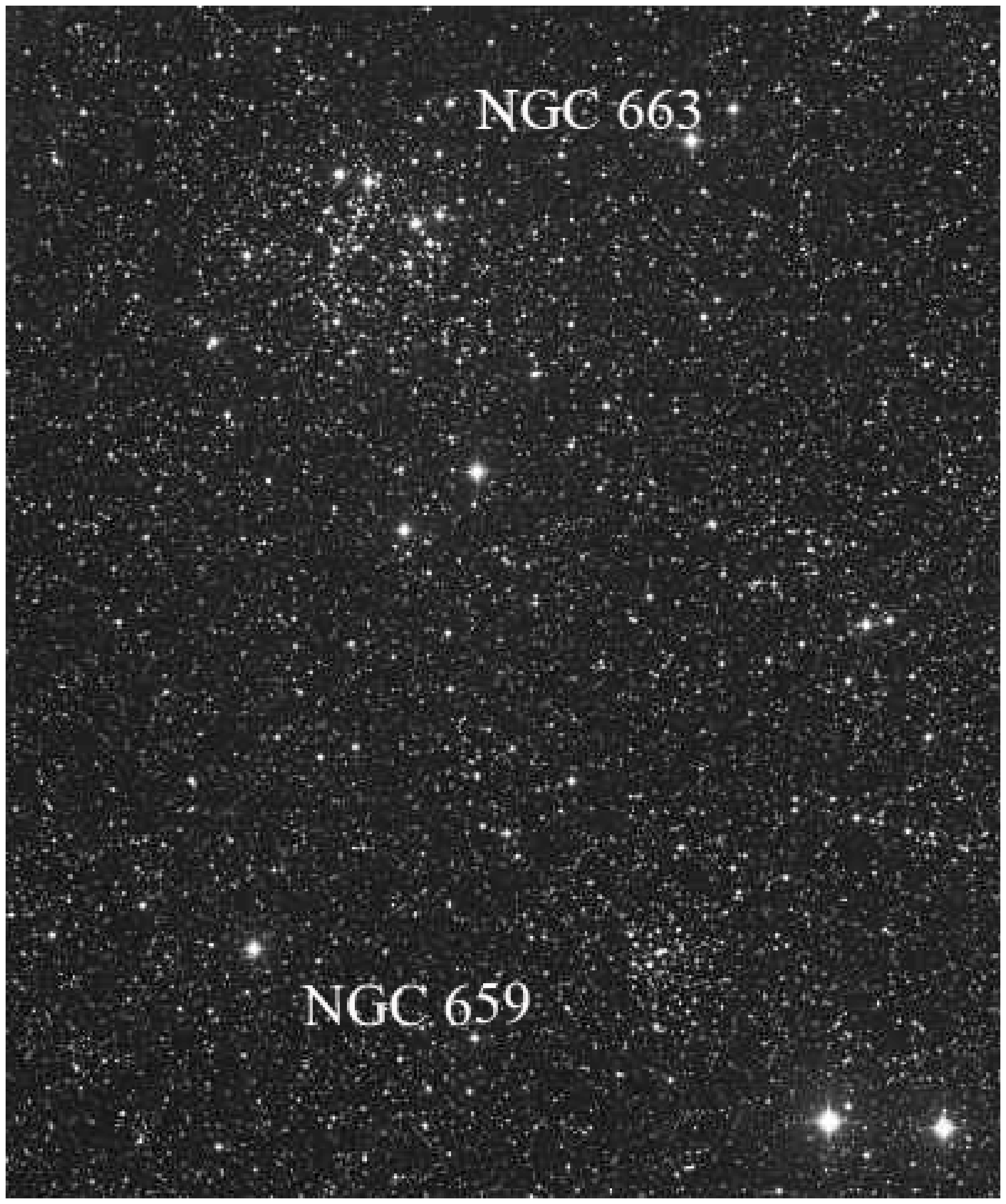} 
        \epsscale{0.302}
        \plotone{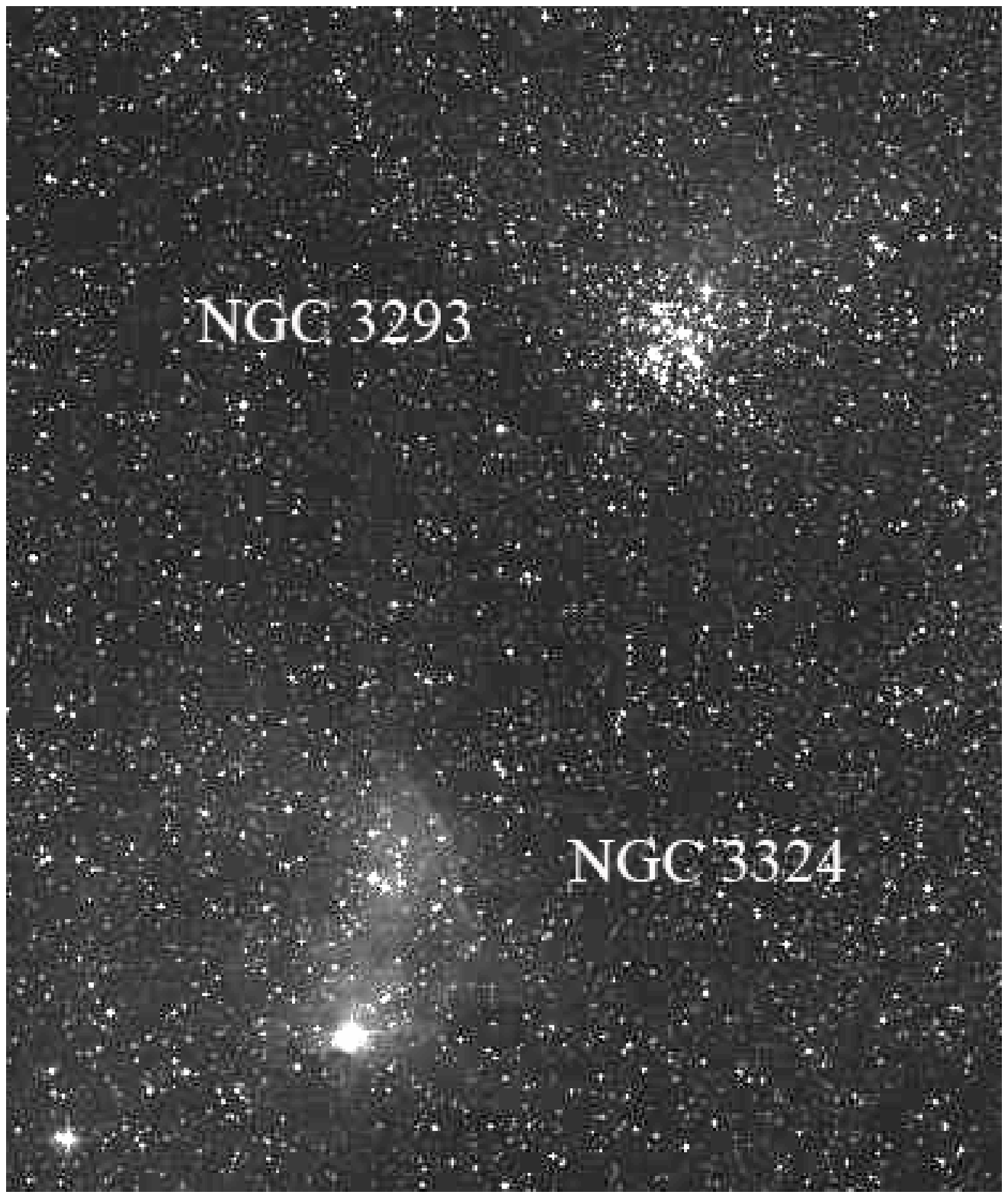} 
        \epsscale{0.285}
        \plotone{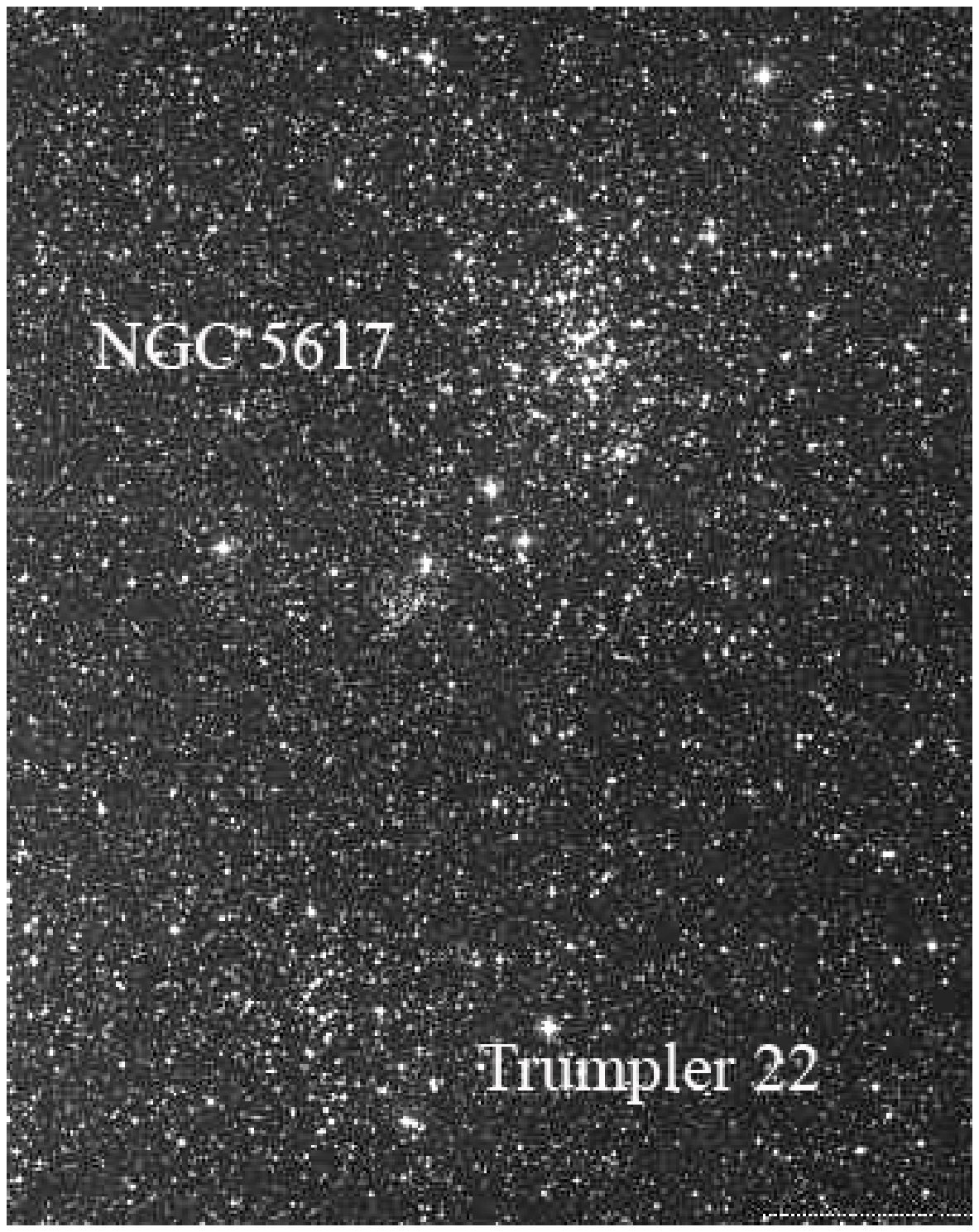} 
        \caption{Examples of real candidate binary clusters cited in the text. 
                 In principle, only the NGC 659/NGC 663 pair (age 16 Myr, separation 25 pc, distance 1938 pc)
                 appears to be following the ionization evolutionary path.
                 Images credit (North is up, East to the left): 
                 (NGC 659/NGC 663) POSSI.E-DSS1 frame, 2.0$\times$1.4 
                 deg$^2$, epoch 1954.75111225188;  
                 (NGC 3293/NGC 3324) POSSI.V-DSS1 frame, 1.0$\times$0.7 
                 deg$^2$, epoch 1987.05060651;  
                 (NGC 5617/Trumpler 22) SERC.J-DSS1 frame, 1.0$\times$0.7 
                 deg$^2$, epoch 1976.1923587345. The NGC 5617/Trumpler 22
                 image also includes the smaller and relatively old (1 Gyr) 
                 open cluster Pismis 19 (or vdBH 160) north from Trumpler 22 
                 and southeast from NGC 5617. This cluster may be interacting 
                 with the other two (de la Fuente Marcos \& de la Fuente Marcos 2009b). 
                 \label{real}}
     \end{figure}

\begin{figure}
        \epsscale{0.9}
        \plotone{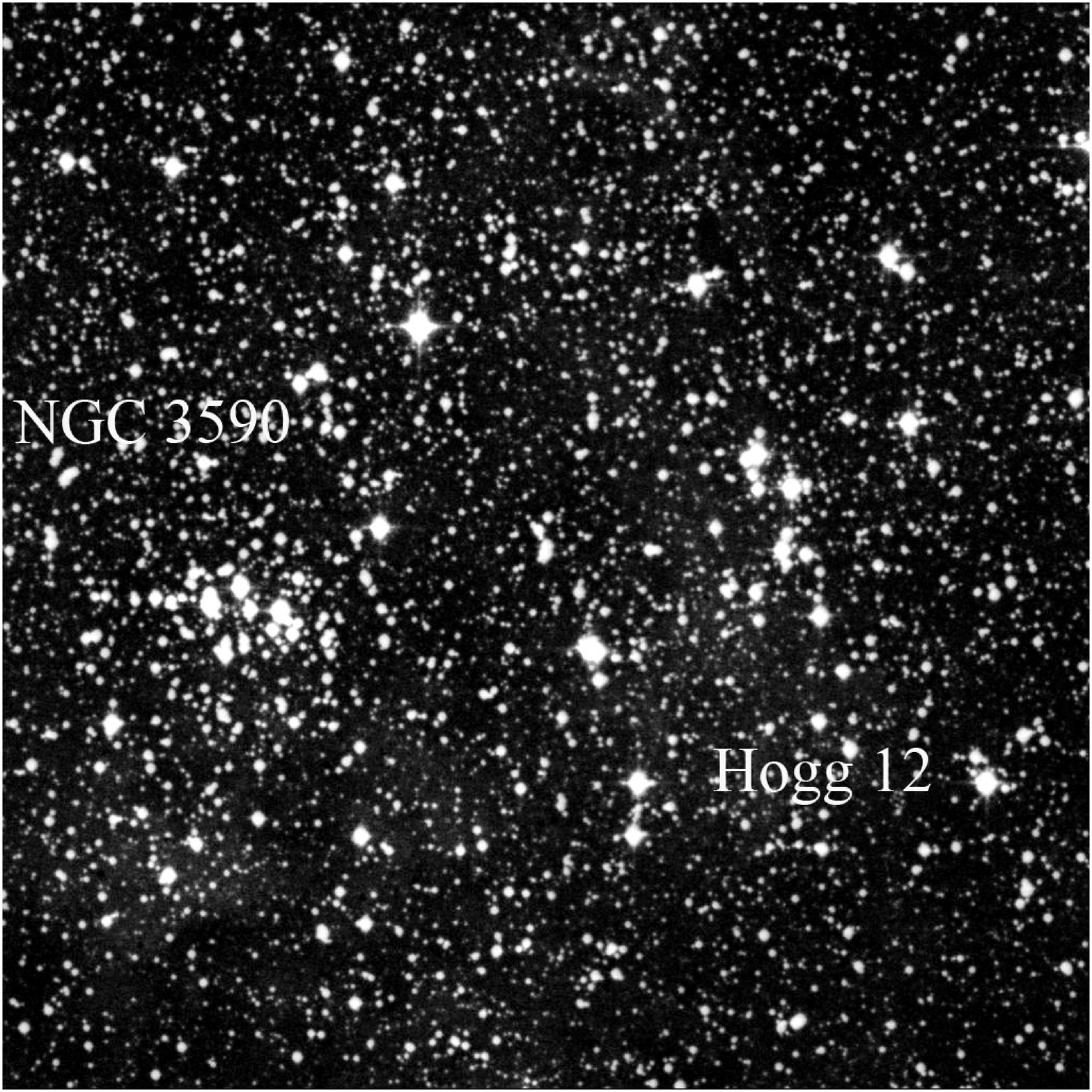} 
        \caption{The NGC 3590/Hogg 12 cluster pair is another strong 
                 candidate to undergo merging. Piatti et al. (2010)
                 have concluded that Hogg 12 is a strongly depleted 
                 but real open cluster and that both open clusters 
                 are located 2 kpc from the Sun. The pair separation
                 is just 3.6 pc and both clusters have similar age 
                 (30 Myr), reddening and metallicity (solar). These
                 authors suggest that it is a strong open cluster
                 binary system candidate. Image credit: ESO.R-MAMA, 
                 0.66 $\mu$m, frame, 11.52$\times$11.52 arcmin$^2$, 
                 epoch 1980.07527720739 (North is up, East to the 
                 left). 
                 \label{real2}}
     \end{figure}

     Figures \ref{evolequal}, \ref{evolhalfsd}, \ref{evolhalfdd} and 
     \ref{evolfourth}, also show probable primordial pairs (age difference $<$30 
     Myr) with pair age $<$ 210 Myr. Following de la Fuente Marcos \& de la Fuente 
     Marcos (2009b), the pair age is assumed to be that of the youngest cluster 
     in the pair. The agreement is very significant. Most pairs, if primordial, 
     appear to be evolving towards the separated twins state. As expected, the observed 
     number of close pairs is consistent with short merging timescale. Only seven pairs 
     in the figures (see also Figure \ref{realmerg}) appear to be in a bound state: 
     NGC 1976/NGC 1981, ASCC 20/ASCC 16, Collinder 197/ASCC 50, NGC 6250/Lynga 14, 
     NGC 3324/NGC 3293, NGC 6613/NGC 6618, and Trumpler 22/NGC 5617. Another candidate 
     (not in the figure) to bound pair is Loden 1171/Loden 1194 (see de la Fuente 
     Marcos \& de la Fuente Marcos 2009b for details). Both Trumpler 22/NGC 5617 and 
     Loden 1171/Loden 1194 seem to be the evolved state of primordial pairs with almost 
     circular initial orbits and original separation $<$ 20 pc. The possible 
     triple cluster NGC 1981/NGC 1976/Collinder 70 is a singular object with 
     the inner pair probably undergoing merging. Some of these binary candidates are
     displayed in Figure \ref{real}. Out of the clusters in this figure, the pairs 
     NGC 3324/NGC 3293 and Trumpler 22/NGC 5617 may be actual bound clusters. In 
     contrast, the pair NGC 659/NGC 663 is likely evolving into the separated 
     twins state. Another example of bound clusters could be the cluster pair 
     NGC 3590/Hogg 12. Piatti et al. (2010) have pointed out this pair as a strong
     open cluster binary system candidate. Both clusters have similar ages (30 Myr),
     reddenings and metallicities. They appear to be located at 2 kpc from the Sun
     and at that distance the pair separation is a mere 3.6 pc. In this case (see 
     Figure \ref{real2}), one of the clusters (Hogg 12) seems to be undergoing 
     extreme tidal distortion. A dramatic example of ongoing merger candidate 
     is the case of the partially embedded massive young cluster NGC 2244 described 
     by Li (2005) where two structures may be separated by just $\sim$ 7 pc.
     Regarding the remarkable double-peaked pair separation distribution observed for 
     young open cluster pairs (de la Fuente Marcos \& de la Fuente Marcos 2009b), the 
     vast majority of observed candidates evolve into separated twins which translates 
     into the dominant peak (25-30 pc) in the pair separation distribution. The secondary 
     peak at smaller separations (10-15 pc) can be explained as due to close pairs 
     in almost circular orbits or undergoing merging. The average tidal radius of open
     clusters at the Solar Circle is nearly 20 pc. Therefore, cluster pairs with
     original separations smaller than the cluster tidal radius are likely to merge
     but those born with wider separations are bound to evolve into the
     separated twins state. Our calculations show that only equal-mass pairs
     formed with originally small separations and nearly circular orbits are likely
     to be observed in close proximity for more than 100 Myr and actually pose as
     genuine binary open clusters. In summary, the bimodal distribution observed in the 
     pair separation histogram is the result of evolutionary effects and not of different 
     formation channels.

     \subsection{How can we identify merger remnants?}
\begin{figure}
        \epsscale{0.6}
        \plotone{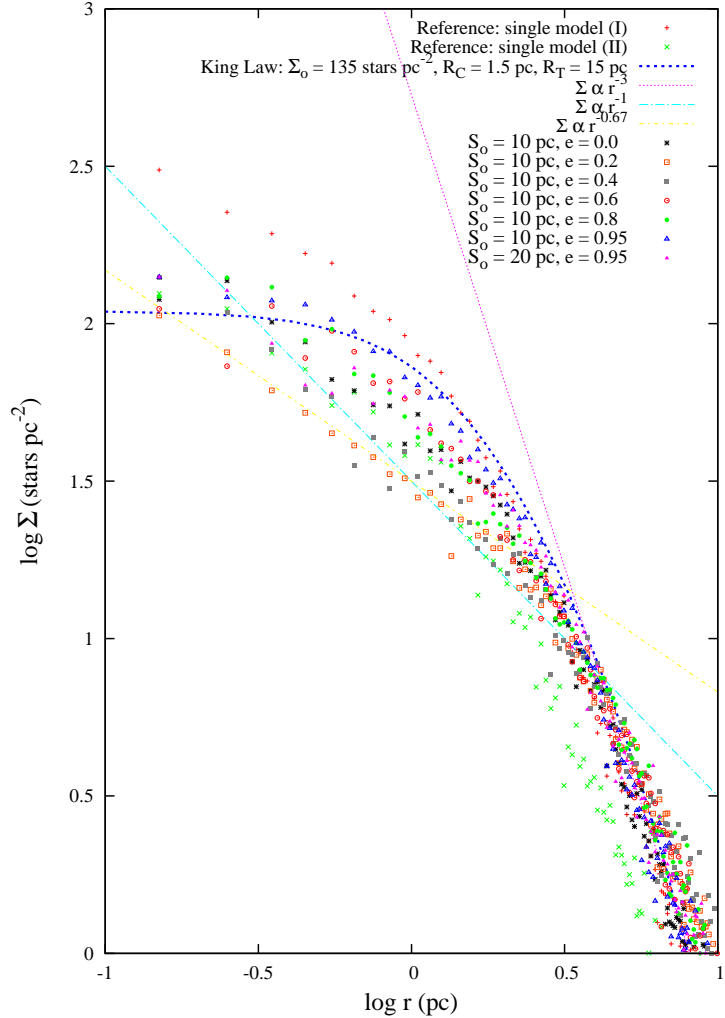} 
        \caption{Radial density profile (top) for merger remnants with $q$ = 1.0
                 at 210 Myr. Two single models with $N$ = 8192 (I), 4096 (II) evolved 
                 in a similar tidal field and a King profile are included for 
                 comparison. The outer regions of merger remnants are 
                 similar to those of an equivalent single cluster but the 
                 number density of the inner regions is a factor 2 lower
                 (150 vs. 300 stars pc$^{-2}$) than that of reference model
                 (I) and close to that of model (II). However, some models show clear 
                 cusps which are absent from single models. The 
                 core of merger remnants from originally eccentric pairs ($e >$ 0.6) 
                 is always denser than that of an equivalent King profile with 
                 similar behavior in the outskirts. Mergers from eccentric pairs
                 take place more rapidly; therefore, the timescale for merging strongly affects
                 the final, observed density profile. The most unusual profiles 
                 are associated to the models with the longest merging timescales 
                 ($e$ = 0.0, 0.2, 0.6). These profiles are $\Sigma(r) \propto r^{-2/3}$
                 in the central regions. Merging of two stellar systems is expected to give 
                 surface density profiles $\Sigma(r) \propto r^{-3}$ (Sugimoto \& Makino 
                 1989; Makino et al. 1990; Okumura et al. 1991). 
                 This also translates into very peculiar velocity profiles (see Figs.
                 \ref{Vrms}, \ref{Vr} and \ref{Vt}). 
                 \label{radialprofile}}
     \end{figure}

     Separated twins are, in principle, easy to identify; just think about the Double 
     Cluster in Perseus (see caption in Figure \ref{separatedtwins}). But, what about 
     merger remnants? One easy to implement test is based on star counts. Figure 
     \ref{radialprofile} shows the surface number density profile for models with 
     $q$ = 1 at 210 Myr. Two single models with $N$ = 8192 and 4096 evolved in a 
     similar tidal field and a King profile (King 1962) are also included as reference. The 
     outer regions of merger remnants are similar to those of an equivalent single cluster 
     having twice the population of the individual clusters but the number density of 
     the inner regions is rather different. In general, merger remnants are expected 
     to be fainter. However, some models ending in merger show clear central cusps 
     which are absent from single models. Merger models characterized by higher 
     central concentration are preferentially associated to pairs with high initial 
     eccentricity and, therefore, shorter merger timescale. They are also the pairs
     in which the mutual tidal disruption has been the weakest as they have been 
     interacting for a shorter period of time. This explains why their cores 
     are nearly 25-50\% denser than that of an equivalent King model. In contrast,
     mergers of low eccentricity pairs have lower central densities. This is the
     result of longer merger timescale and enhanced mutual tidal disruption.
     This lower density translates into higher production 
     of hierarchical systems in the form of temporarily stable triple and quadruple 
     stellar systems with respect to single models. This is to be expected as higher 
     stellar density increases the probability of relatively close-range gravitational 
     interactions that quickly destroy multiple systems. In general, 
     the central regions of merger remnants are distinctively different from those of 
     clusters evolving without companions but haloes look very similar. 

\begin{figure}
        \epsscale{0.8}
        \plotone{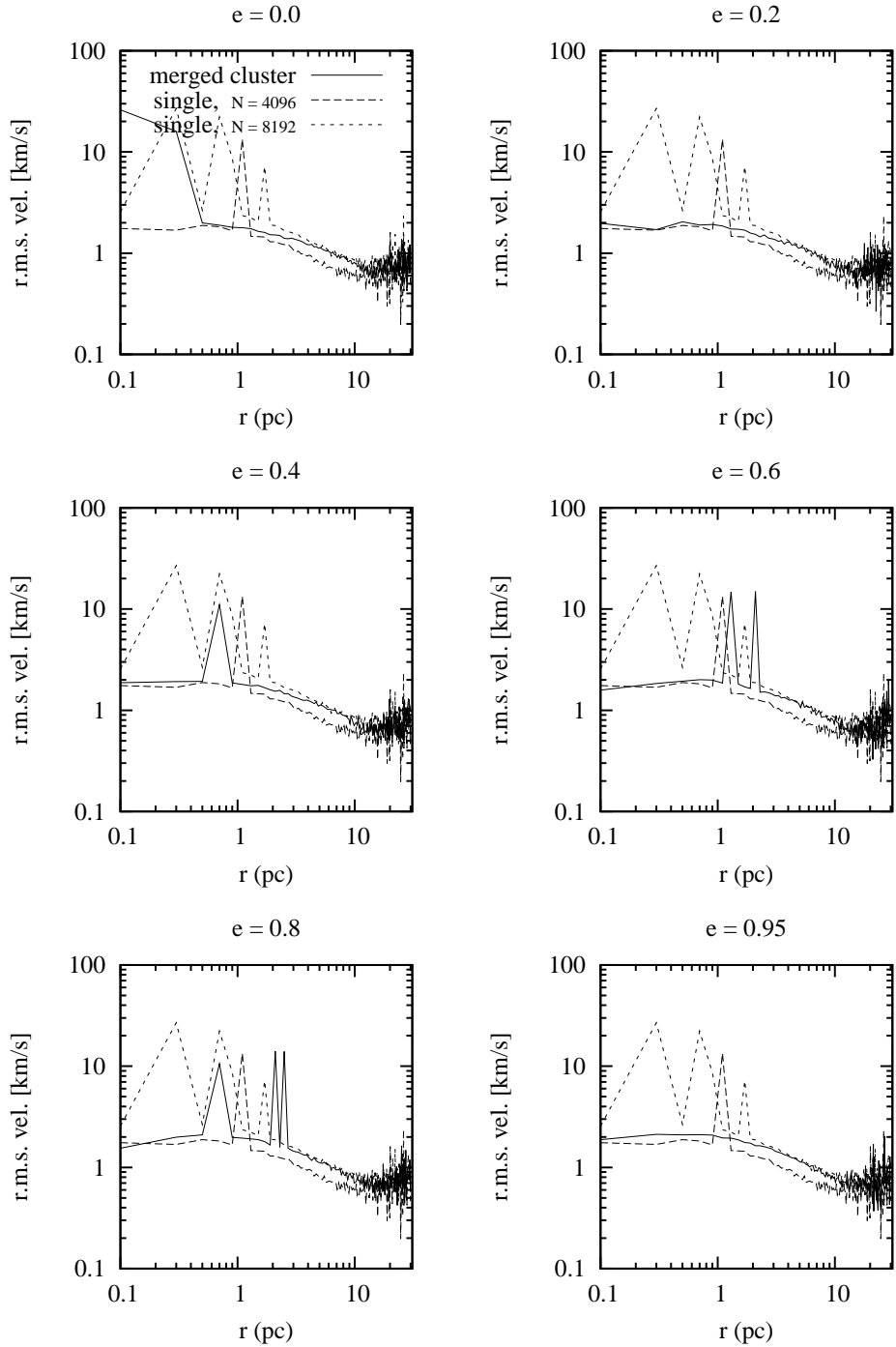} 
        \caption{Root mean square velocity profile for the same merged and single 
                 clusters in Fig. \ref{radialprofile}, after 210 Myr. The rms value 
                 is always greater than or equal to the average as it includes
                 the standard deviation as well. In general, 
                 the rms velocity of the central regions of merger remnants is 
                 significantly lower than that of an equivalent single cluster 
                 with $N$ equal to twice the population of the merged clusters but 
                 just slightly higher than that of a cluster with the same population 
                 of the individual clusters. The outer regions are more similar to
                 those of the larger $N$ cluster. Velocities are referred to the cluster or
                 merger center of masses (CM). 
                 \label{Vrms}}
     \end{figure}

\begin{figure}
        \epsscale{0.8}
        \plotone{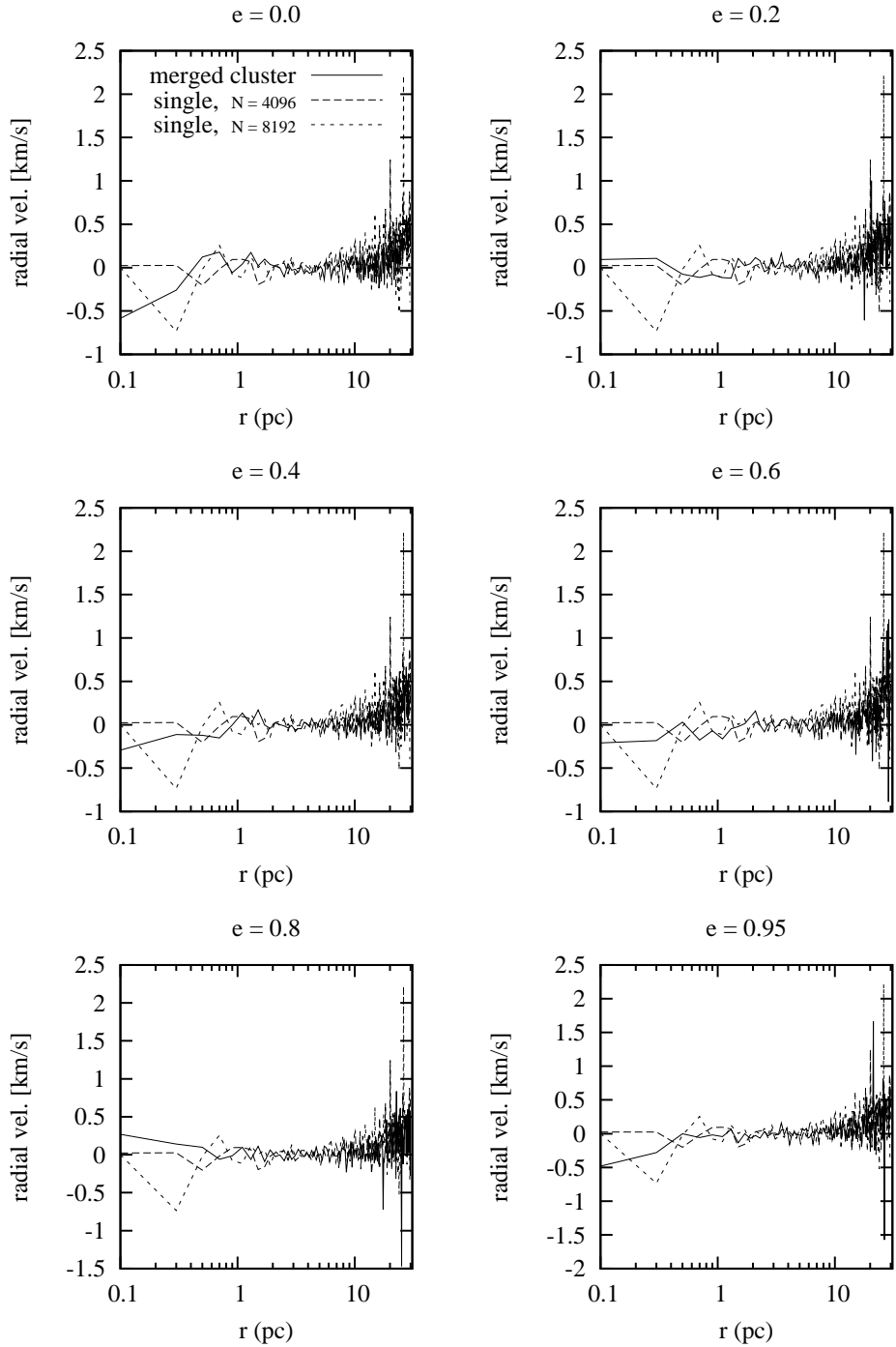} 
        \caption{Radial velocity profile for the same merged and single
                 clusters in Figs. \ref{radialprofile} and \ref{Vrms}, after 210 Myr. 
                 In general, the average radial velocity of the central regions of 
                 merged clusters is lower than that of the single model with 
                 $N$ = 4096. The standard deviation is however nearly 40\% the
                 average value; therefore, the profiles are consistent. 
                 Velocities are referred to the cluster or merger CM.
                 \label{Vr}}
     \end{figure}

\begin{figure}
        \epsscale{0.8}
        \plotone{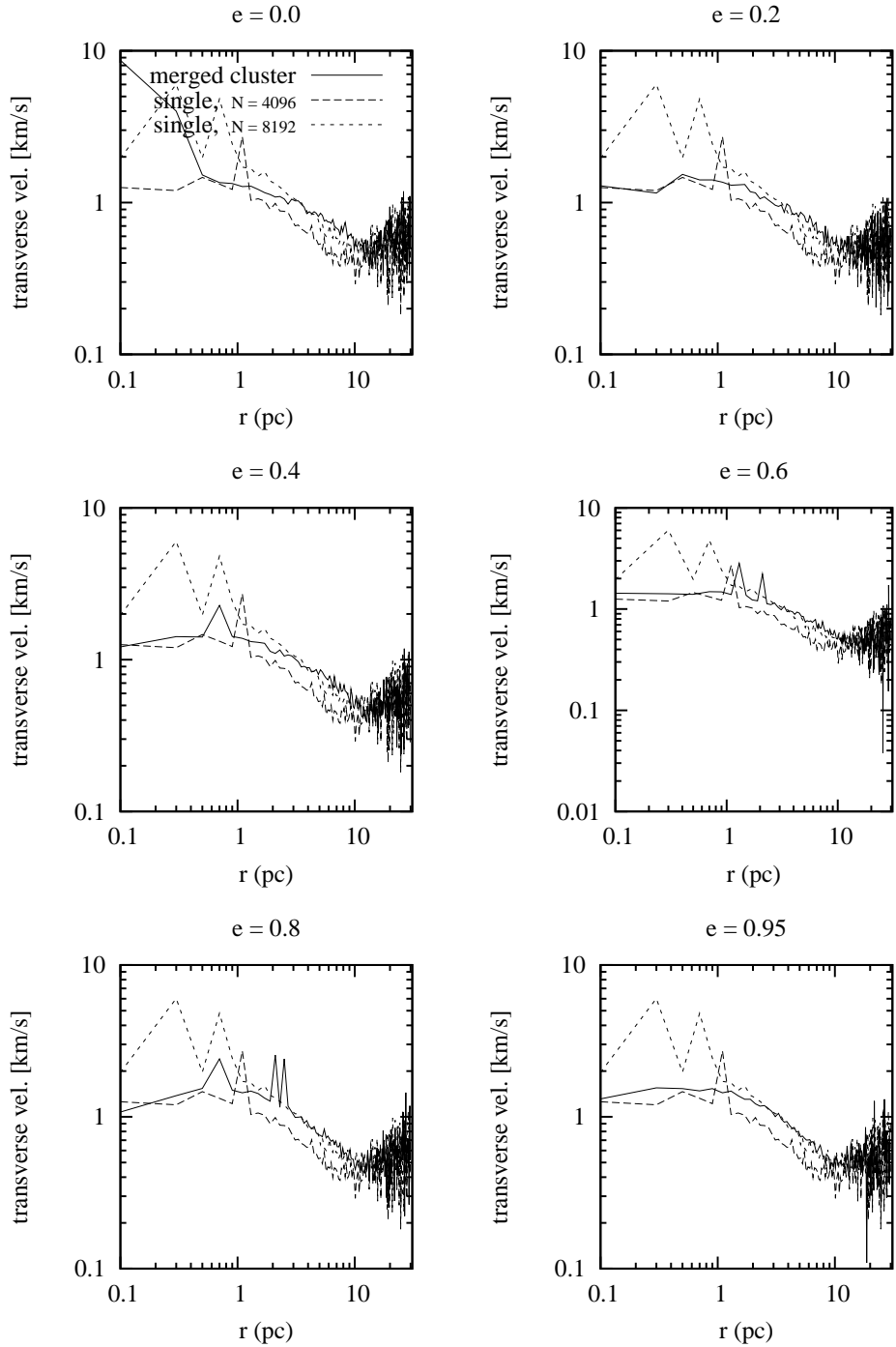} 
        \caption{Transverse velocity profile for the same merged and single
                 clusters in Figs. \ref{radialprofile}, \ref{Vrms} and \ref{Vr}, 
                 after 210 Myr. In general, transverse velocity profiles for mergers 
                 are very smooth. In the halo of the cluster they match that of an equivalent 
                 single cluster with $N$ equal to twice the population of the pre-merger 
                 cluster. However, in the inner regions the average transverse velocity 
                 is significantly lower, similar to that of a cluster with the same 
                 population of the individual pre-merger clusters. Beyond 10 pc from the center
                 profiles are very similar. The model with the longest merging timescale ($e$ = 0.0)
                 exhibits the most unusual behavior with very strong rotation at the center.
                 Velocities are referred to the cluster or 
                 merger CM.
                 \label{Vt}}
     \end{figure}

     Regarding the kinematic signature of merging, it is also obvious in the central 
     regions. Data in the figures discussed here are referred to the center of masses 
     of the merged/single cluster. In this analysis we use the root mean square velocity
     defined as the square root of the average velocity-squared of the stars in the
     cluster, their radial velocity and transverse velocity. The $V_{rms}$ is always 
     greater than or equal to the average as it includes the standard deviation 
     ($V^{2}_{rms} = <V>^{2} + \sigma^2$). The radial velocity is given by
     \begin{equation}
        V_r = \frac{\vec{v} \cdot \vec{r}}{|\vec{r}|} \,,
     \end{equation}
     where $\vec{r}$ is the position of the star and $\vec{v}$ its velocity. Finally, 
     the transverse velocity is defined by 
     \begin{equation}
        V_t = \frac{|\vec{r} \times \vec{v}|}{|\vec{r}|} \,.
     \end{equation}
     The transverse velocity is also the product of the angular speed $\omega$ and the
     magnitude of the position vector. Therefore and for a given star cluster region, 
     transverse velocity and angular speed exhibit the same behavior: both increase or 
     decrease concurrently. The modulus of the specific angular momentum is 
     the magnitude of the position vector times the transverse velocity.  
     Figure \ref{Vrms} shows the rms velocity for concentric shells similar to those in 
     Figure \ref{radialprofile} for merged models and the two reference single models 
     already presented. The value has been summed over all the stars in the shell considered
     and divided by the number of particles inside that shell.
     In general, the outcome of a typical merged model shows a value of the rms 
     velocity in the core intermediate to those of the reference models and it gets closer
     to that of the most massive reference model in the outer parts of the cluster. The 
     behavior beyond the half-mass or 
     effective radius is similar for all the models, the rms velocity decreases 
     gradually outwards. It is, however, not so clear how this peculiar velocity 
     distribution can be used to identify merger remnants. The behavior of the
     radial velocity profile in Figure \ref{Vr} is, however, easier to interpret.
     In mergers, radial velocities tend to be lower or even negative in the central
     regions of the cluster (see Fig. 3, top panel in Baumgardt et al. 2003). This
     trend is observed in several of the studied cases and it may point out to collapse
     of the central regions induced by the gravogyro instability (see below) or to
     statistical fluctuations. The transverse
     velocity profile is displayed in Figure \ref{Vt}. In general and up to 10 pc
     from the center of the merger these profiles are very smooth. The value of the 
     transverse velocity in the central pc of the merger remains remarkably
     constant in clear contrast with what is observed for single cluster models.
     This can be interpreted as evidence for a moderate amount of global rotation.
     On theoretical grounds (King 1961), rotation causes a slight increase in the
     rate at which a cluster is losing stars. This is consistent with the
     relatively smaller final population observed in merged models.
     
     \subsection{Mergers, synchronization and gravogyro instabilities}
     Single self-gravitating and globally rotating $N$-body systems are affected by a
     little known instability that is playing a role here: the gravogyro instability.
     Initially proposed by Inagaki \& Hachisu (1978) and Hachisu (1979, 1982) and
     further studied by Akiyama \& Sugimoto (1989) using numerical techniques, the 
     gravogyro instability or gravogyro catastrophe is triggered when specific angular 
     momentum is removed (by escaping stars) from the cluster and a deficit in the 
     supporting centrifugal force is induced. In other words, the escape of stars from
     the core to the outer regions transports angular momentum from the inner regions of 
     the cluster to the outer parts. As a result, the inner regions react to compensate
     the loss of centrifugal force contracting in order to increase rotation. Faster
     rotation induces additional mass loss. The gravogyro effect increases the average
     angular speed of the central regions of the cluster decreasing the average value 
     of the local angular momentum. The overall result is faster dynamical evolution
     of the star cluster and higher mass loss. The gravogyro instability leads to further 
     mass loss through the galactic tidal boundary. This process causes the contraction of 
     the core of the star cluster as observed in Figure \ref{radialprofile} (models with
     $e_o$ = 0.2 and 0.6, mainly). 

     But before the onset of the gravogyro instability, a merger must take place in order 
     to have a rotating system. This is the result of yet another instability: the 
     synchronization instability. The numerical/theoretical work of Sugimoto \& Makino 
     (1989) and Sugimoto et al. (1991) showed that binary star clusters undergo an 
     instability at a critical separation and this process is able to trigger rapid merging of 
     star clusters. The orbit of the bound cluster pair circularizes and shrinks due
     to the loss of angular momentum carried away by escaping particles. As the separation 
     of the binary system decreases, the orbital period of the binary becomes shorter and 
     the mutual tidal effects stronger. Concurrently, the spin of each cluster increases 
     following the orbital motion due to the induced tidal torque attempting to synchronize 
     cluster rotation and orbital period. This transformation of orbital angular momentum into 
     rotational angular momentum leads to an instability at a critical separation when the 
     exchanged orbital angular momentum becomes insufficient to supply the necessary angular 
     momentum for spin synchronization. At that moment merging takes place as the orbital 
     angular momentum is no longer capable of balancing the gravitational attraction between 
     the two star clusters. Figures \ref{evolequal}, \ref{evolhalfsd} and \ref{evolhalfdd} 
     clearly indicate that in our models and after the pair separation 
     becomes $<$ 2 pc, merging proceeds very quickly. The gyration radius of a self-gravitating
     system of stars can to first-order be approximated by $R_{vir}$ (e.g., Portegies
     Zwart \& Rusli 2007); therefore, merger occurs when the intercluster separation 
     falls below twice the gyration radius. This process is similar to the tidal locking effect observed in 
     planetary systems. The main difference here is that in planet-satellite interactions 
     both conversion of orbital angular momentum into spin and vice versa are possible. On 
     the other hand and in absence of dissipation (drag forces), the outcome of tidal 
     locking is always a stable binary configuration not a merger as in the case of star 
     clusters. The main reason for the lack of stable outcome in the case of star clusters 
     is that the total angular momentum is not actually preserved as stellar evolution, 
     rotation-induced escapees (those resulting from the gravogyro effect), and two-body 
     relaxation are concurrently and constantly removing angular momentum from the system. 
     The total angular momentum removal rate depends on many factors, the age of the cluster 
     being one of them. Other facts to consider here are the masses of the clusters, the mass 
     ratio, and the galactic tidal field.
     
     The gravogyro and synchronization instabilities are somehow related as they involve
     rotation and transfer of angular momentum but they are not the same process. For 
     originally non-rotating, merging clusters, synchronization occurs first and then, 
     eventually and after global cluster rotation is induced, the gravogyro instability 
     starts on the resulting merger. The final phases of the synchronization instability 
     and the onset of the gravogyro instability overlap; in fact, and as described above,
     the enhanced escape rate caused by the gravogyro instability helps to deplete the
     remaining orbital angular momentum. The most dramatic example is observed for the
     slowest merging model, the one with $e_o$ = 0.0. 

\begin{figure}
        \epsscale{0.8}
        \plotone{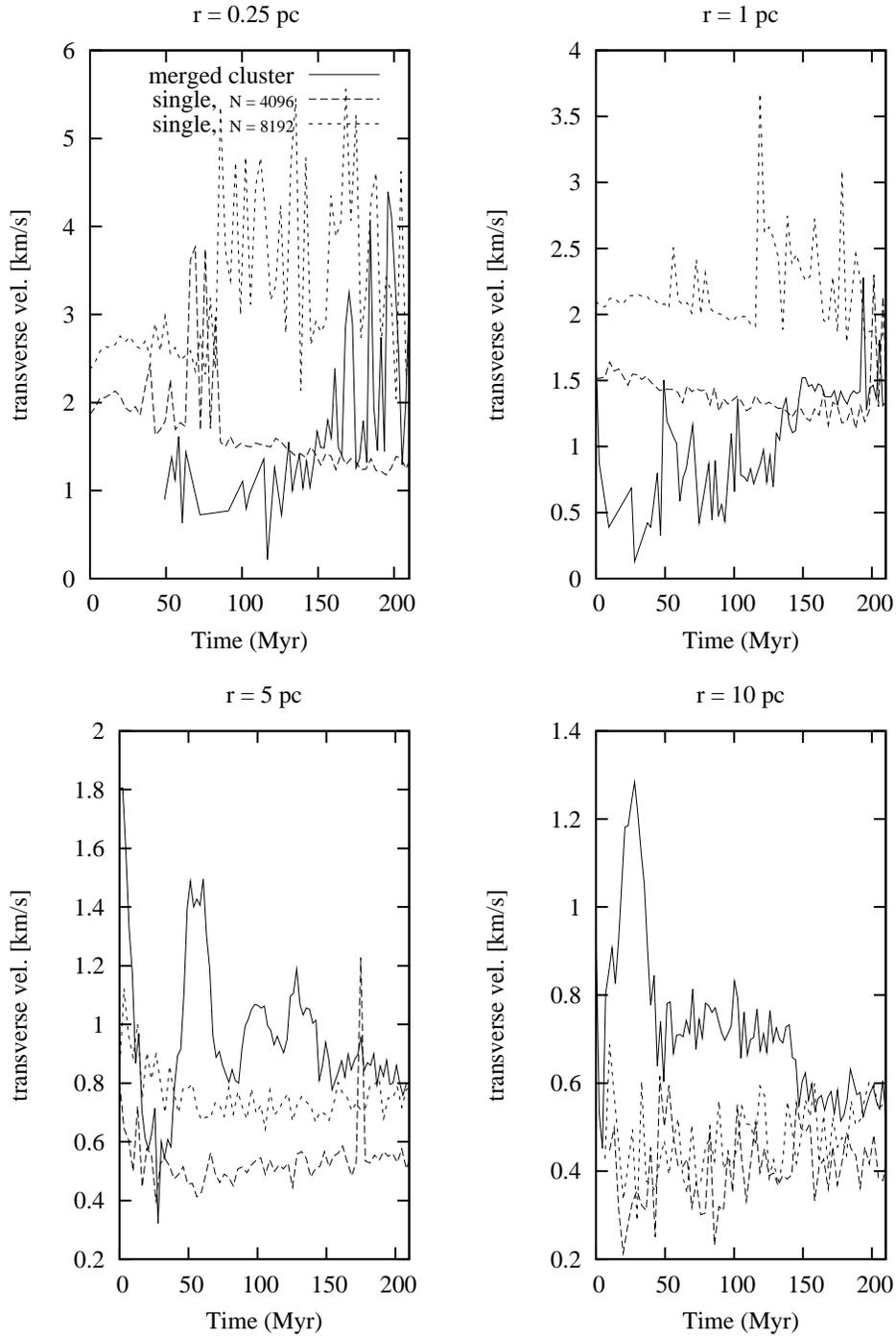} 
        \caption{Evolution of the transverse velocity over time for representative
                 shells of the model displayed in Figs. \ref{merging} and 
                 \ref{trajectory}. Shells are $r \in$ [0, 0.25], $r \in$ [0.75, 1.0],
                 $r \in$ [4.75, 5.0], and $r \in$ [9.75, 10.0]. After merging ($t \sim$
                 150 Myr), the average transverse velocity increases over time in
                 the central regions which is the typical signature of the gravogyro 
                 instability. On the other hand, it tends to decrease slightly or 
                 remain constant in the outer regions. Fluctuations (standard deviation)
                 around the average value in this plot and the following three amount
                 to nearly 40\% (error bars are not displayed for clarity).
                 \label{Vtevol}}
     \end{figure}

\begin{figure}
        \epsscale{0.8}
        \plotone{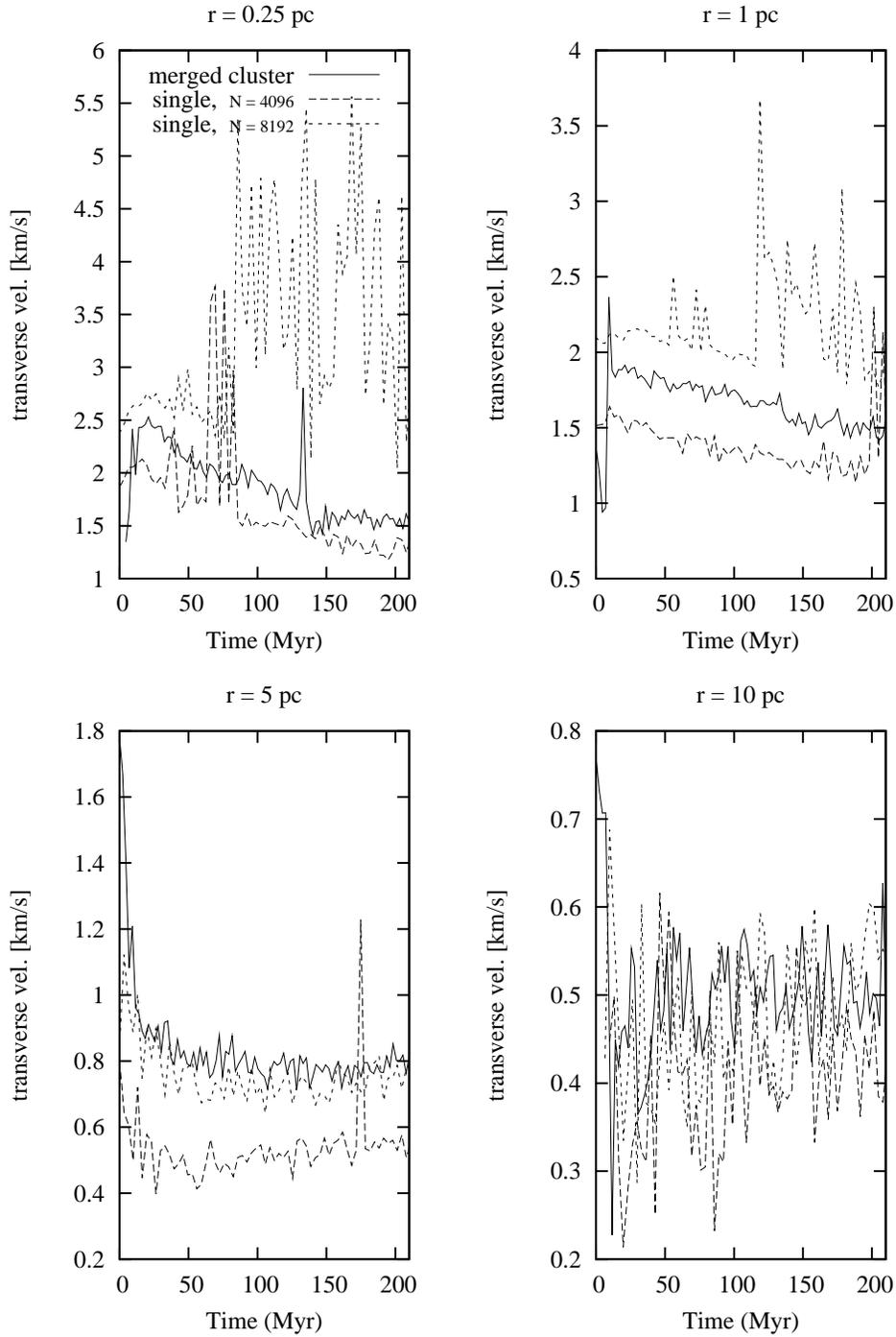} 
        \caption{Same as Fig. \ref{Vtevol} but for the model with $q$ = 1.0, 
                 $S_o$ = 10.0 pc and $e_o$ = 0.95. Here the long term evolution 
                 is similar to that of the single models with $N$ = 4096. The
                 average transverse velocity decreases over time and approaches
                 a constant value. The gravogyro instability is now absent as
                 expected for a rotating multicomponent model with concurrent 
                 energy equipartition at work.
                 \label{Vtevol2}}
     \end{figure}

\begin{figure}
        \epsscale{0.8}
        \plotone{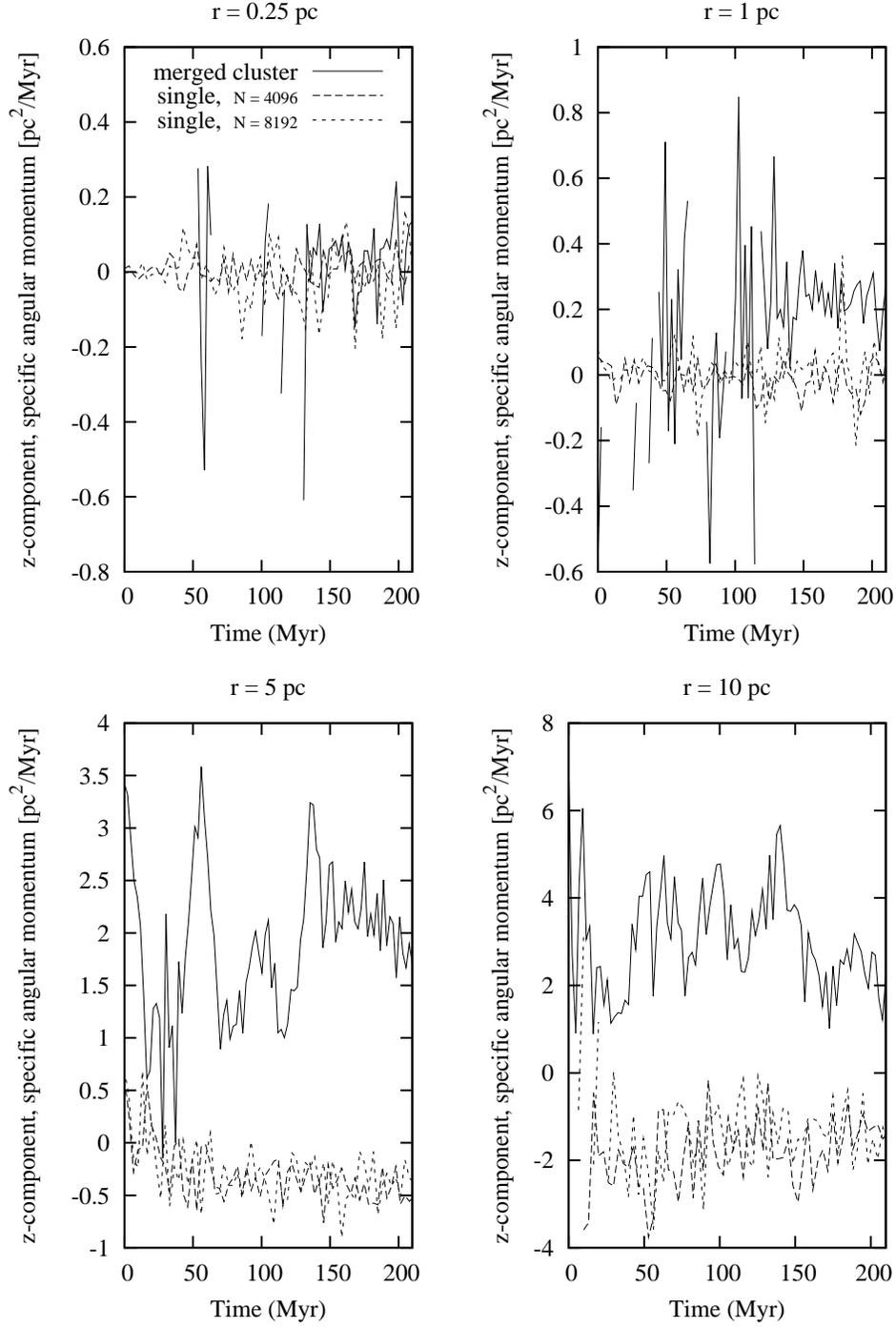} 
        \caption{Time evolution of the $z$-component of the specific angular
                 momentum of the model in Figs. \ref{merging} and
                 \ref{trajectory}. After merging, the remnant exhibits
                 slight decrease of angular momentum.
                 \label{jzevol}}
     \end{figure}

\begin{figure}
        \epsscale{0.8}
        \plotone{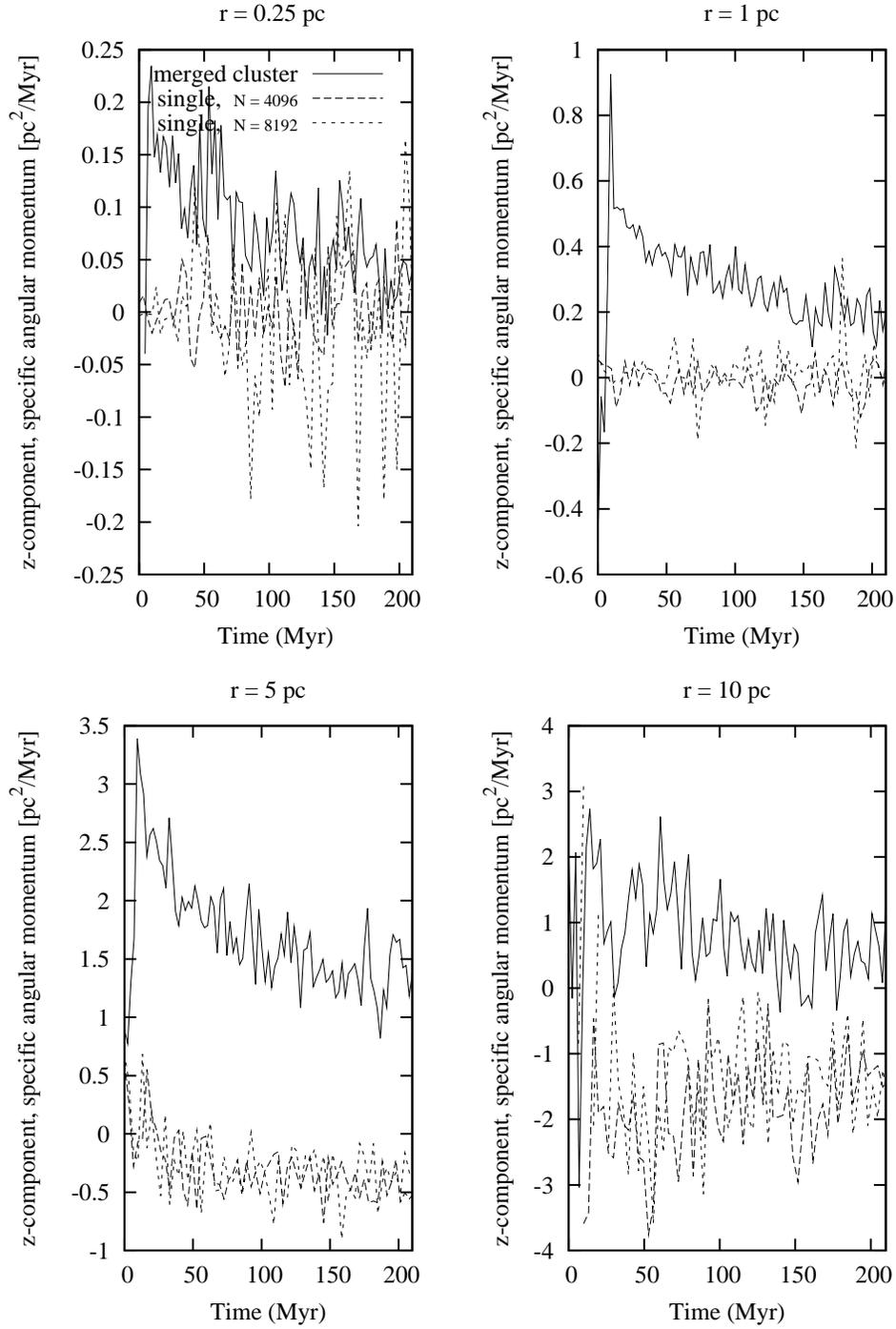} 
        \caption{Same as Fig. \ref{jzevol} but for the model with $q$ = 1.0,
                 $S_o$ = 10.0 pc and $e_o$ = 0.95. The evolution is now
                 completely different. 
                 \label{jzevol2}}
     \end{figure}

     Single star clusters formed out of unstable disks might exhibit some degree of
     primordial rotation. Ernst et al. (2007) studied $N$-body models of rotating 
     globular clusters to confirm that the gravogyro instability takes place in
     monocomponent clusters (clusters with equal-mass stars). They found that the
     $z$-component of the specific angular momentum decreases over time for the
     inner regions of the clusters increasing, in return, the average angular speed 
     of the same areas (and therefore the transverse velocity); see their Figs. 3 
     and 4. For these models the
     effect is very important with a significant amount of angular momentum being
     transfered outwards. The induced deficit of centrifugal force triggers the
     collapse of the core of the star cluster. If the Galactic tidal field is 
     included in the calculations, rotation increases the escape rate dramatically.
     In sharp contrast, their models of systems with two-mass components (two mass 
     groups for a very simple IMF) show that the effect of rotation is rather negligible. 
     In this case and 
     if rotation is primordial, two concurrent processes are at work: mass segregation 
     and rotation. Both compete to accelerate the collapse of the cluster core but mass 
     segregation clearly dominates. In summary, they found that if energy equipartition 
     is at work the role of rotation is somewhat secondary but in absence of energy 
     equipartition, rotation through the gravogyro effect speeds up and controls the
     evolution of the cluster. This is what we observe in our models. Figures 
     \ref{Vtevol} and \ref{Vtevol2} show the evolution of the average value of the
     transverse velocity for relevant concentric cluster shells (see the caption
     for details) of two representative models: the ones with the longest and shortest,
     respectively, merging timescale. By the time ($\sim$ 150 Myr) the 
     clusters in the slowest merging model actually merge, energy equipartition has been 
     achieved and the gravogyro effect is the strongest. Conversely, the fastest merging 
     model shows that rotation is quickly lost from the merger remnant through 
     energy equipartition; the time evolution of the average transverse velocity being 
     similar to that of a single cluster model with $N$ = 4096. As for the time evolution
     of the $z$-component of the specific angular momentum in our models, representative 
     results are displayed in Figs. \ref{jzevol} and \ref{jzevol2}. For the slowest 
     merging model and after merging, systematic loss of angular momentum is observed 
     with the outer regions sporting the largest share of angular momentum. In the case 
     of an early merger, angular momentum also decreases over time but the magnitude of 
     this angular momentum is much higher and positive as expected for a rotating system.

     \subsection{How do these results compare with those from previous works?}
     Unfortunately, little attention has been devoted to the topic of binary cluster 
     evolution. In a ground breaking paper, Sugimoto \& Makino (1989) used $N$-body
     simulations to study the merging timescale of two identical star clusters
     ($N$ = 2048). They found that the interaction provoked circularization of the
     orbits of the clusters and synchronization of the spin of each cluster with
     its orbital revolution. The loss of orbital angular momentum caused the eventual
     demise of the pair. The entire process was described by them 
     as the synchronization instability. This phenomenon is also responsible for
     merging in our models. Their work was continued in Makino et al. (1991) for
     non-equal clusters. Merging of two stellar systems usually gives surface
     density profiles $\Sigma(r) \propto r^{-3}$ (Sugimoto \& Makino 1989; Makino et al.
     1990; Okumura et al. 1991). In our models this is only true for the outskirts
     of merger remnants; the central regions can be better described by 
     $\Sigma(r) \propto r^{-1}$ or $r^{-2/3}$ (see Figure \ref{radialprofile}).
     Using analytical arguments, Ballabh \& Alladin (2000) showed that 
     merging always occurs if the distance of closest approach is about twice the 
     sum of the dynamical radii of the clusters. This is confirmed by our calculations.
     De Oliveira et al. (2002) used
     $N$-body simulations to study the dynamical status of the cluster pair
     NGC 1912/NGC 1907. Their simulations found the formation of stellar bridges
     similar to the ones found in our calculations. Faster encounters produce weaker
     tidal debris in the bridge area. Our simulations are also consistent with this result.

     The most realistic simulations of binary cluster evolution so far 
     have been performed by Portegies Zwart \& Rusli (2007) using the {\small kira} integrator 
     of the {\small starlab} simulation environment. Their calculations were aimed at 
     understanding the nature of the cluster pair NGC 2136/NGC 2137, in the Large 
     Magellanic Cloud. Their models are, therefore, more massive; their primary 
     clusters have $N$ in the range [9,000, 24,855] with smaller secondaries moving in 
     circular orbits. They found that cluster pairs with initial separation 
     smaller than 15-20 pc merge in $<$60 Myr. Pairs with larger initial 
     separation tend to become even more widely separated over time. In spite of 
     the different code and initial conditions used, our results are, in general, 
     fully compatible with theirs. There is, however, a major difference induced by the
     fact that they do not take into account the background galactic tidal field. They
     did not observe the extreme tidal distortion displayed in Figure \ref{spaghetti}. 
     Besides, they only follow their models for about 100 Myr. Their main models have $q$ = 0.167; 
     this value triggers catastrophic destruction in our models. The soft ionization widely found 
     in our simulations was originally described in a series of little known but very 
     interesting papers (Theis 2001, 2002a,b). Within the context of globular 
     cluster formation, Theis' simulations found that identical twin clusters may 
     merge or evolve into well separated twins sharing a common galactic orbit.  

     Our calculations clearly show that the impact of merging on the evolution
     of close and therefore young open cluster pairs is all but negligible. The 
     same can likely be said about young star cluster pairs in other galaxies. 
     On the other hand, the rapid decrease in star cluster numbers for ages 
     older than 20 Myr ({\it infant mortality}) observed in our Galaxy and 
     others and usually attributed exclusively to the catastrophic gas ejection 
     mechanism originally proposed by Hills (1980) can also be the result of 
     merging or tidal disruption in close primordial pairs. Merging and tidal
     disruption of the less massive companion may easily halve (at least) the
     initial population of relatively close open cluster pairs. Merging, disruption 
     and infant mortality, concurrently, can efficiently reduce the number of 
     observed young star clusters and accelerate dramatically the transition of stars
     born in clusters to the field populations. 

     It may be argued that the loss of clusters 
     through merging can only be significant if the fraction of binary clusters 
     at birth is high, but is the primordial semi-major distance distribution 
     supportive of this? If the Orion Nebula star forming complex could be 
     considered as representative of the kind of environment in which most open 
     clusters actually form, the answer may be in the affirmative. Current
     available evidence (de la Fuente Marcos \& de la Fuente Marcos 2009b) 
     suggests that the group NGC 1981/NGC 1976/ Collinder 70/$\sigma$ Ori may 
     form several close pairs, all of them with separations $<$ 30 pc and 
     few Myr old. There is
     also a statistical tool that may help to understand the timeline of the 
     relative importance of these processes: the generalized fractal dimension 
     or multifractal analysis. De la Fuente Marcos \& de la Fuente Marcos (2006) 
     showed that the generalized fractal dimension changes very significantly for 
     clusters younger than 40 Myr but remains almost constant for older clusters. 
     Catastrophic gas ejection is an essentially self-driven process and operates 
     on all scales; merging or extreme tidal distortion only operates on small 
     length-scales within star-forming complexes where the typical intercluster 
     distance is $<$ 30 pc. Processes operating on all scales keep the value of 
     the generalized fractal dimension almost constant across the multifractal 
     spectrum. The opposite is observed when processes operate only for objects 
     in close proximity. Our numerical results are consistent with the fractal 
     dimension results if most pairs are born in originally very eccentric orbits 
     ($e >$ 0.5) and/or with very different masses. Those are the ones merging 
     (or being tidally destroyed) within 40 Myr of forming.

  \section{Conclusions}
     Using available observational evidence, de la Fuente Marcos \& de la Fuente 
     Marcos (2009b) have demonstrated that the population of binary open clusters is 
     statistically significant and that the fraction of candidate binary clusters 
     in the Milky Way disk is comparable to that in the Magellanic Clouds, $\sim$
     10\%. Out of this population, nearly 40\% of them can be classified as genuine 
     primordial binary open clusters although only a relatively small fraction 
     ($\sim$17\%) appear to be able to survive as conspicuous pairs for more than 
     25 Myr. The distribution of open cluster separations exhibits an apparent
     peak at 10-15 pc, analogous to the one observed in both LMC and SMC. 
     Here, we have used $N$-body simulations in an attempt to understand how 
     primordial binary open clusters evolve and what initial orbital elements are
     required in order to explain their observed properties. V\'azquez et al. (2010) 
     concluded that available observational evidence indicates that double cluster 
     lifetimes are short and that is what simulations confirm. Our main conclusions 
     can be summarized as follows.
     
     \begin{enumerate}

        \item Long term stability of binary open clusters appears not to be 
              possible. Primordial binary open clusters seem to be inherently 
              transient objects, at least for the range of cluster parameters
              explored here. 

        \item The results of our simulations interpreted within the context of 
              the available observational data clearly indicate that the vast 
              majority of primordial binary open clusters gradually evolve into 
              well separated objects. After one Galactic rotation the separation 
              is in the range 200-500 pc but they exhibit relatively prominent 
              tidal bridges. Formation of separated twins is driven by mass loss 
              induced by stellar evolution, two-body relaxation, and mutual 
              tidal disruption.

        \item Close primordial binary open clusters quickly merge into a single
              object. The merging timescale depends on the orbital and physical 
              characteristics of the pair. Most close pairs merge within 100 Myr
              of formation. For eccentric pairs the merging scale is even
              shorter, about 50 Myr. Our numerical results and the short 
              characteristic observational survival time for candidate primordial 
              pairs, 25 Myr, strongly suggest that nearly 80\% of primordial 
              binary open clusters are born in systems with orbital eccentricities 
              $>$ 0.5. Merging is driven by the synchronization instability.

        \item For clusters pairs of appreciably different masses extreme tidal
              distortion or shredding of the secondary is observed, 
              causing the complete disruption of the less massive cluster within 
              one Galactic orbit. Remnants of shredded clusters may show up
              in kinematic studies even if they cannot be detected as stellar 
              overdensities. 

        \item The gravogyro instability shapes the spatial and kinematic structure 
              of merger remnants but the effect is only dominant for primordial
              pairs in almost circular orbits. 

        \item The observed candidate pair separation histogram shows a bimodal
              distribution. In the light of the present results, the two peaks are 
              mainly the result of ionization and merging. They do not appear to
              be the result of different formation channels but different evolutionary
              paths.

        \item The sharp decline in open cluster numbers observed for objects 
              older than about 20 Myr can be explained by three different 
              processes operating concurrently: catastrophic gas expulsion, tidal disruption 
              and merging. This can efficiently reduce the number of observed young star 
              clusters and accelerate dramatically the transition of stars
              born in clusters to the field populations. 
     
     \end{enumerate}

     It has often been assumed that binary open cluster formation in the Milky Way 
     is uncommon. In contrast, our results indicate that the lives of primordial
     binary clusters are violent and hazardous. Close pairs (if formed) 
     merge in a short timescale, being the shortest for very eccentric pairs, 
     secondaries in low mass ratio pairs are rapidly destroyed, and wide pairs 
     quickly ionize in the background tidal field due to mass loss. As open clusters 
     are actually born in close proximity (complexes), these appear to be the genuine 
     reasons behind the apparent lack of binary open clusters in our Galaxy. Star
     cluster binarity is, therefore, a transient phenomenon. 

     The effects and trends observed in the present set of simulations are robust for
     the range of open cluster parameters studied. As usual, it is potentially
     dangerous to make unwarranted extrapolations to larger/smaller or denser clusters.
     It could be possible that for much larger and denser star clusters the merging 
     timescale is longer. Nevertheless, the absence of binary globular clusters in the 
     Milky Way appears to indicate that long-term binary cluster stability is, in fact,
     unlikely. The role of the gravogyro effect of the evolution of merger remnants
     appears to be well documented in our present work but larger simulations are
     needed to better understand the statistical strength of this process. 

  \acknowledgments
     The authors would like to thank S. Aarseth and K. Nitadori for providing 
     the code used in this research. The authors also would like to acknowledge 
     the help of J. L. Mazo Fr\'{\i}as, N. Pedone, and P. Chan of Dell Computer 
     with the T5500+S1070 system. In preparation of this paper, we made use of 
     the NASA Astrophysics Data System and the astro-ph e-print server. This 
     research has made use of the WEBDA database operated at the Institute of 
     Astronomy of the University of Vienna, Austria. This work also made use of 
     the ALADIN, SIMBAD and VIZIER databases, operated at the CDS, Strasbourg, 
     France.

%
%

\clearpage
%
%

%
%

\clearpage
%
%

%
%

\clearpage
%
%

%
%

\clearpage
%
%

%
%

\clearpage
%
%

%
%

\clearpage
%
%

%
%

\clearpage
%
%

%
%

\clearpage
%
%

%
%

\clearpage
%
%

%
%

\clearpage
%
%

%
%

\clearpage
%
%

%
%

\clearpage
%
%

%
%
\clearpage
%
%

%
%
\clearpage
%
%

%
%
\clearpage
%
%

%
%
\clearpage
%
%

%
%
\clearpage
%
%

%
%
\clearpage
%
%

%
%
\clearpage
%
%

%
%
\clearpage
%
%

\clearpage

         \begin{table}
          \fontsize{8} {12pt}\selectfont
          \tabcolsep 0.10truecm
          \caption{Results of the 72 computations after 210 Myr}
          \begin{tabular}{ccccccc}
           \hline
           \multicolumn{1}{c}{$\setminus e_o$} &
           \multicolumn{1}{c}{0.00}                      &
           \multicolumn{1}{c}{0.20}                      &
           \multicolumn{1}{c}{0.40}                      &
           \multicolumn{1}{c}{0.60}                      &
           \multicolumn{1}{c}{0.80}                      &
           \multicolumn{1}{c}{0.95}                      \\
       $S_o$(pc) & (Myr)/(pc)   & (Myr)/(pc)   & (Myr)/(pc)   & (Myr)/(pc)   & (Myr)/(pc)   & (Myr)/(pc)   \\   
                 & (Myr)        & (Myr)        & (Myr)        & (Myr)        & (Myr)        & (Myr)        \\   
           \hline
             \multicolumn{7}{c}{$q$ = 1.0}              \\   
             \multicolumn{7}{c}{$N_1$ = $N_2$ = 4096, $M_{1, 2} = 2048 M_{\odot}$, $R_{vir}$ = 1 pc, $R_p$ = 0.59 pc, 
                                $R_T$ = 18 pc}   \\   
           \hline
             10  & 175/0.44$^*$ & 95/0.21      & 72/0.22      & 39/0.16      & 32/0.23      & 37/0.14      \\   
                 & 46           & 35           & 28           & 23           & 19           & 17           \\
             20  & $\infty$/369 & $\infty$/343 & $\infty$/325 & $\infty$/299 & $\infty$/263 & 100/0.19     \\   
                 & 130          & 99           & 78           & 64           & 54           & 48           \\
             30  & $\infty$/497 & $\infty$/485 & $\infty$/470 & $\infty$/448 & $\infty$/421 & $\infty$/391 \\   
                 & 239          & 182          & 144          & 118          & 99           & 88           \\
           \hline
             \multicolumn{7}{c}{$q$ = 0.5}              \\
             \multicolumn{7}{c}{same density}              \\
             \multicolumn{7}{c}{$N_1$ = 4096, $N_2$ = 2048, $M_1 = 2048 M_{\odot}$, $M_2 = 1024 M_{\odot}$,
                                $R_{vir 1}$ = 1 pc, $R_{vir 2}$ = 0.79 pc,} \\
             \multicolumn{7}{c}{$R_{p 1}$ = 0.59 pc, $R_{p 2}$ = 0.46 pc, $R_{T}$ = 18 pc}   \\   
           \hline
             10  & 78/0.18      & 64/0.10      & 53/0.55      & 61/0.22      & 59/0.32      & 51/0.17      \\   
                 & 53           & 40           & 32           & 26           & 22           & 19           \\
             20  & $\infty$/332 & $\infty$/313 & $\infty$/321 & $\infty$/272 & $\infty$/246 & $\infty$/230 \\   
                 & 150          & 114          & 91           & 74           & 62           & 55           \\
             30  & $\infty$/459 & $\infty$/445 & $\infty$/437 & $\infty$/421 & $\infty$/394 & $\infty$/384 \\   
                 & 276          & 210          & 166          & 136          & 114          & 101          \\
           \hline
             \multicolumn{7}{c}{$q$ = 0.5}              \\
             \multicolumn{7}{c}{different density}              \\   
             \multicolumn{7}{c}{$N_1$ = 4096, $N_2$ = 2048, $M_1 = 2048 M_{\odot}$, $M_2 = 1024 M_{\odot}$,
                                $R_{vir 1}$ = 1 pc, $R_{vir 2}$ = 0.50 pc,} \\
             \multicolumn{7}{c}{$R_{p 1}$ = 0.59 pc, $R_{p 2}$ = 0.29 pc, $R_{T}$ = 18 pc}   \\   
           \hline
             10  & 78/0.52      & 43/0.38      & 67/0.46      & 56/0.50      & 64/0.40      & 64/0.24      \\   
                 & 53           & 40           & 32           & 26           & 22           & 19           \\
             20  & $\infty$/301 & $\infty$/308 & $\infty$/317 & $\infty$/273 & $\infty$/254 & 159/0.48     \\   
                 & 150          & 114          & 91           & 74           & 62           & 55           \\
             30  & $\infty$/443 & $\infty$/428 & $\infty$/429 & $\infty$/404 & $\infty$/384 & $\infty$/370 \\   
                 & 276          & 210          & 166          & 136          & 114          & 101          \\
           \hline
             \multicolumn{7}{c}{$q$ = 0.25}              \\
             \multicolumn{7}{c}{tidal disruption}              \\
             \multicolumn{7}{c}{$N_1$ = 4096, $N_2$ = 1024, $M_1 = 2048 M_{\odot}$, $M_2 = 512 M_{\odot}$,
                                $R_{vir 1}$ = 1 pc, $R_{vir 2}$ = 0.63 pc,} \\
             \multicolumn{7}{c}{$R_{p 1}$ = 0.59 pc, $R_{p 2}$ = 0.37 pc, $R_{T}$ = 18 pc}   \\   
           \hline
             10  & $\infty$/134$^{\star}$ & $\infty$/116 & $\infty$/96  & $\infty$/91  & $\infty$/77  & $\infty$/46  \\   
                 & 58           & 44           & 35           & 29           & 24           & 21           \\
             20  & $\infty$/282 & $\infty$/276 & $\infty$/262 & $\infty$/233 & $\infty$/212 & $\infty$/178 \\   
                 & 164          & 125          & 99           & 81           & 68           & 60           \\
             30  & $\infty$/415 & $\infty$/401 & $\infty$/405 & $\infty$/380 & $\infty$/358 & $\infty$/351 \\   
                 & 302          & 230          & 182          & 149          & 125          & 111          \\
           \hline
          \end{tabular}
          \begin{list}{}{}
            {\footnotesize
              \item[*] Merging timescale in Myr / cluster pair spatial separation in pc, 
              \item[]  $P_{orb}$ in Myr (Equation \ref{period})
              \item[$\star$]  For models with $q$ = 0.25 the separation between the primary cluster and the disrupted
                              secondary is quoted. 
             \item[] $S_o$ is the initial value of the apoclustron distance (see the text for details).
             \item[] $e_o$ is the initial value of the eccentricity.
             \item[] $q = M_2 / M_1$.
            }
          \end{list}
          \label{results}
         \end{table}

\end{document}